\documentclass[useAMS,usenatbib]{mn2e}
\usepackage{graphicx} 
\usepackage{times}
\usepackage{epsfig}
\usepackage{rotating}
\usepackage{sidecap}
\usepackage{longtable}
\usepackage{lscape}
\def\lesssim{\mathrel{\hbox{\rlap{\hbox{\lower4pt\hbox{$\sim$}}}\hbox{$<$}}}}
\let\la=\lesssim
\def\gtrsim{\mathrel{\hbox{\rlap{\hbox{\lower4pt\hbox{$\sim$}}}\hbox{$>$}}}}
\let\ga=\gtrsim

\title[LFQPOs of GRO J1655-40]{Discovery of two simultaneous non-harmonically related Quasi-Periodic Oscillations in the 2005 outburst of the black-hole binary GRO J1655-40}
\author[S. Motta et al.]{S.~Motta$^{1}$, J. Homan$^{2}$, T. Mu\~noz-Darias$^{3}$, P. Casella$^{4}$, T.M. Belloni$^{1}$, B. Hiemstra$^{5}$, M. M\'endez$^{5}$\\
$^{1}$INAF-Osservatorio Astronomico di Brera, Via E. Bianchi 46, I-23807 Merate (LC), Italy\\
$^{2}$MIT Kavli Institute for Astrophysics and Space Research, 70 Vassar Street, Cambridge, MA 02139, USA\\
$^{3}$School of Physics and Astronomy, University of Southampton, Southampton, Hampshire, SO17 1BJ, United Kingdom\\
$^{4}$INAF, Osservatorio Astronomico di Roma, Via Frascati 33, I-00040, Monteporzio Catone, Italy\\
$^{5}$ Kapteyn Astronomical Institute, University of Groningen, PO Box 800, 9700 AV Groningen, the Netherlands\\
}

\begin{document}
\maketitle
\begin{abstract}

We studied the low-frequency quasi–-periodic oscillations (LFQPOs) in the black hole GRO J1655-40 during the 2005 outburst, using data from the Rossi X-ray Timing Explorer. All LFQPOs could be identified as either type B or type C using previously proposed classification schemes. In the soft state of the outburst the type-C LFQPOs reached frequencies that are among the highest ever seen for LFQPOs in black holes. At the peak of the outburst, in the ultra-luminous state, the power spectrum showed two simultaneous, non-harmonically related peaks which we identified as a type-B and a type-C QPO. The simultaneous presence of a type-C and type-B QPO shows that at least two of the three known LFQPO types are intrinsically different and likely the result of distinct physical mechanisms. We also studied the properties of a broad peaked noise component in the power spectra of the ultra-luminous state. This noise component becomes more coherent with count rate and there are strong suggestions that it evolves into a type-B QPO at the highest observed count rates.

\end{abstract}


\section{Introduction}

Fast time variability is an important characteristic of black hole X-ray binaries (BHXBs) and a key ingredient in understanding the physical processes in these systems. BHXBs show a variety of X-ray spectral/variability states, representing different accretion regimes (see \citealt{Belloni2010} for a recent review). These states are often easy to identify in hardness-intensity diagrams (HID, see \citealt{Homan2001}). \citealt{Munoz-Darias2011} recently showed that the overall strength of the rapid variability, as measured by the rms (\textit{root main square deviation}), is a good tracer of these states as well. In an rms-intensity diagram (RID, see \citealt{Munoz-Darias2011}) the spectral/variability states  become apparent without relying on spectral information, and the transitions between them are often better defined than in the HID. Fast (aperiodic) variability is generally studied through the inspection of power density spectra (PDS; see van der Klis 1989). Most of the power spectral components in the PDS of BHXBs are broad and can take the form of a wide power distribution over several decades of frequency or of a more localized peak (quasi-periodic oscillations, QPOs). %BHXBs PDS often show dramatic changes in shape on very different time-scales. In particular QPOs display both secular variations and fast dramatic changes  (see \citealt{Belloni2011} for a review). 

QPOs have been detected in many BHXBs and are thought to originate in the innermost regions of the accretion flow around the black hole. 
%QPOs in neutron star X-ray binaries were originally discovered by EXOSAT (\citealt{Stella1986}), while 
Low-frequency QPOs (LFQPOs), with frequencies ranging from a few mHz to $\sim$10 Hz, are common features in BHXBs. They were first observed with \emph{Ariel 6} in GX 339-4 (\citealt{Motch1983}) and observations with \emph{Ginga} provided the first indications for the existence of multiple types of LFQPOs (see e.g. \citealt{Miyamoto1991} for the case of GX 339-4 and \citealt{Takizawa1997} for GS 1124-68). Observations performed with the Rossi X-ray Timing Explorer (RXTE) have led to an extraordinary progress in our knowledge on the properties of the LFQPOs in BHXBs (see \citealt{VDK2006}, \citealt{Remillard2006}, \citealt{Belloni2011} for recent reviews). Three main types of LFQPOs are currently recognized: types A, B, and C. They were originally identified in the PDS of XTE J1550-564 (see \citealt{Wijnands1999}; \citealt{Homan2001}; \citealt{Remillard2002}), and have now been seen in several other sources (see, e.g., \citealt{Casella2004} and \citealt{Motta2011a}).  

Despite  the fact that LFQPOs have been known for several decades, their origin is still not understood, and there is no consensus about their physical nature, to the point that it is still not clear if the different types of QPOs share a common origin or arise from different physical phenomena\footnote{For theoretical models on the origin of QPOs, see e.g. \citealt{Esin1997}, \citealt{Titarchuk1999}, \citealt{Tagger1999}, \citealt{Done2007}, \citealt{Ingram2011} and references therein.}. However, the study of LFQPOs provides an indirect way to explore the inner accretion flow around black holes (and neutron stars). In particular, their association with specific spectral states and their phenomenology suggest that they could be a key ingredient in understanding the physical conditions that give origin to the different states.

Recently, \cite{Motta2011a} showed that the differences between the three types of LFQPOs extend beyond the properties that led to the definition of the ABC classification. In particular, they showed that at least two of the three types showed a very different dependence of frequency on power-law flux, suggesting that these QPOs (type B and C) might be the result of different physical processes.
%In this paper we further study the physical meaning of the ABC classification, in order to verify the hypothesis that the different LFQPOs are indeed intrinsically different phenomena and we report on a result that favors this hypothesis. 
Up to now different types of LFQPOs have never been detected simultaneously, hence the final proof of an intrinsic difference between different types of LFQPOs was missing. Here we present evidence of a simultaneous detection of two different types of QPOs in the PDS of the black hole binary GRO J1655-40. 

%----------------------------------------------------------------------------------
\begin{figure}
\begin{center}
\includegraphics[width=8.5cm]{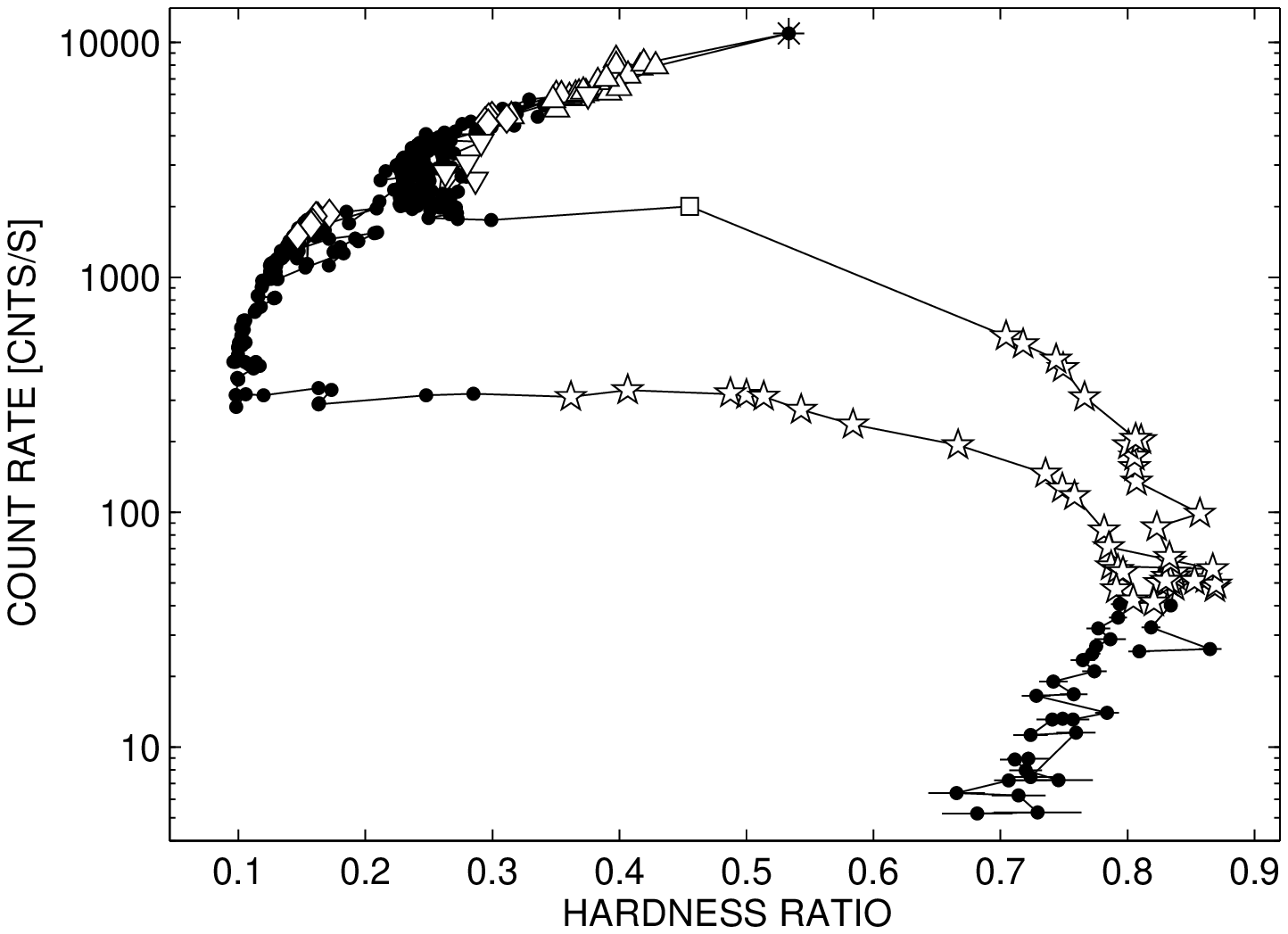}
\includegraphics[width=8.5cm]{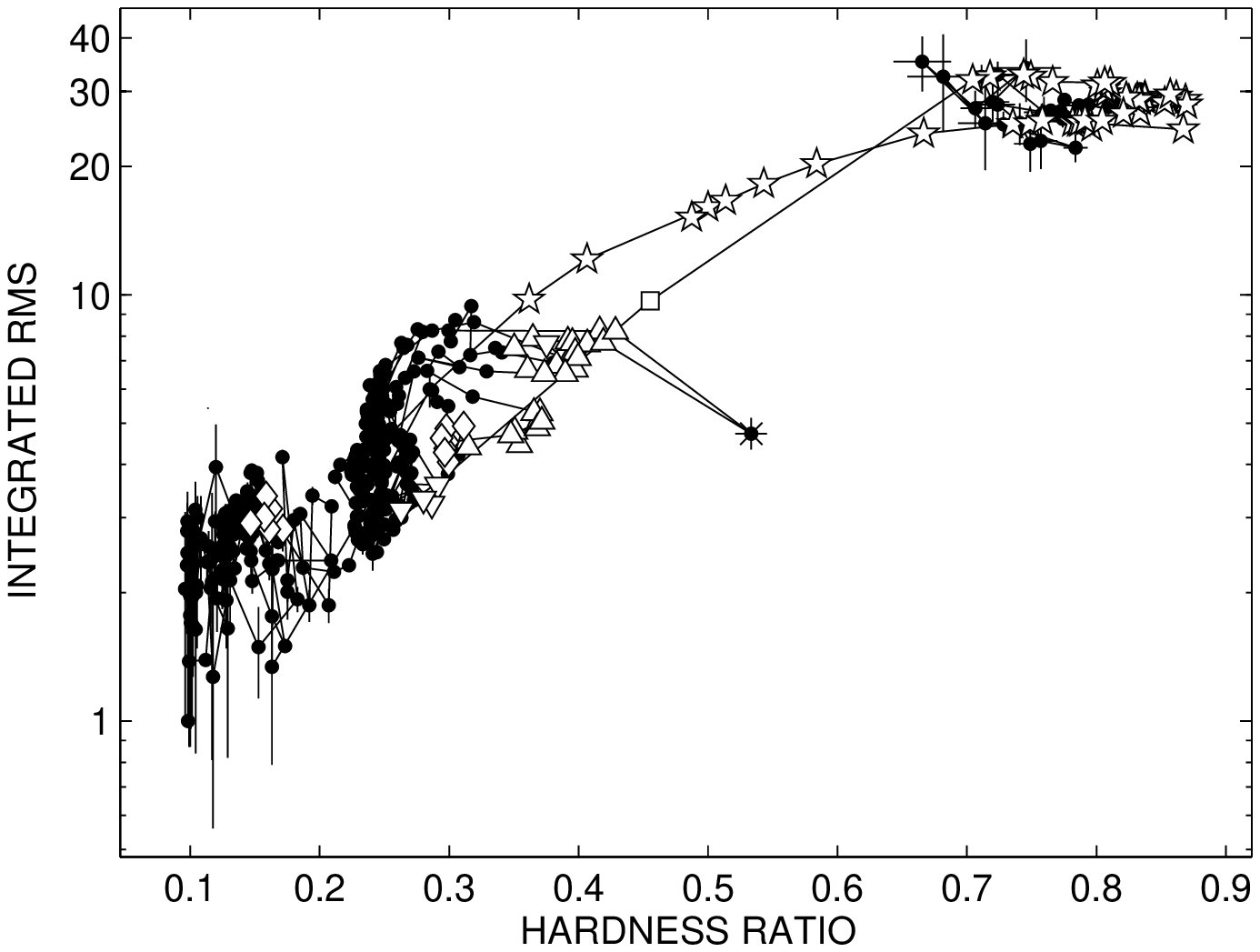}
\includegraphics[width=8.5cm]{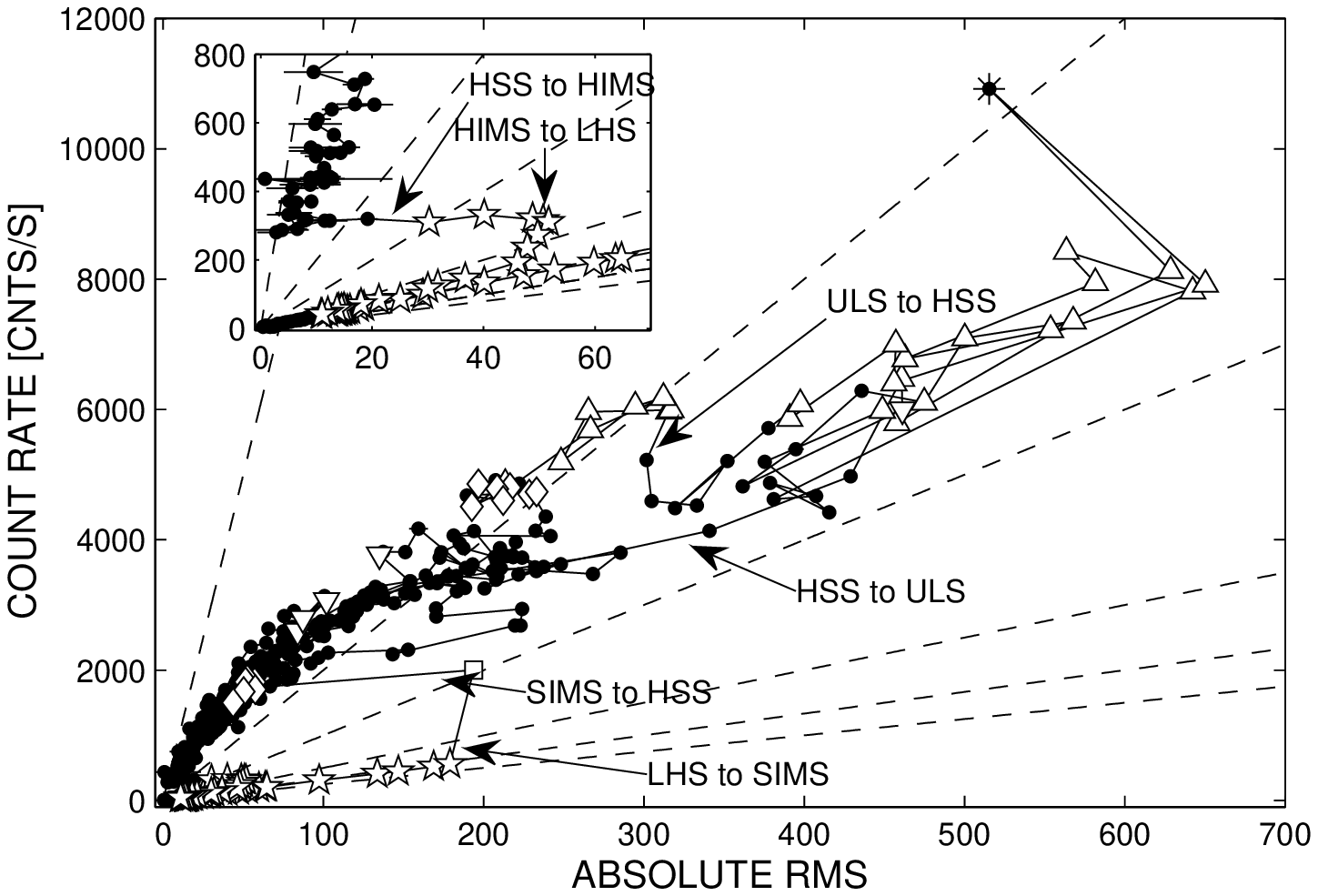}
\caption{Upper panel: hardness-Intensity diagram. Middle panel: hardness-rms diagram. Bottom panel: rms-intensity diagram. The dashed lines represent (from left to right) the 1, 5, 10, 20, 30, 40 per cent fractional rms levels. The inset shows a closeup view of the hardening phase of the outburst. Arrows indicate the state transitions described in Sec. \ref{sec:results}. Notice that the transitions to and from the ULS have been identified through a detailed timing analysis and through the inspection of the HID/RID. The source starts its outburst in the bottom-left corner and follows the diagram in an anti-clockwise way. For further details on the RID, see \citealt{Munoz-Darias2011}.
 In all the panels the solid line joins observations as a function of the time. Different symbols mark the PDS type described in the text (see Sec. \ref{sec:timing}) and shown in Fig. \ref{fig:PDS}. Stars: type 1; upward triangles: type 2; diamonds: type 3; downward triangles: type 4; square: type 5; asterisk: type 6. Black circles mark observations where no QPO were observed. See Sec. \ref{sec:observations} for information about the diagrams. }\label{fig:RID}
\end{center}
\end{figure}
%----------------------------------------------------------------------------------

%----------------------------------------------------------------------------------
\begin{figure}
\begin{center}
\includegraphics[width=8.6cm]{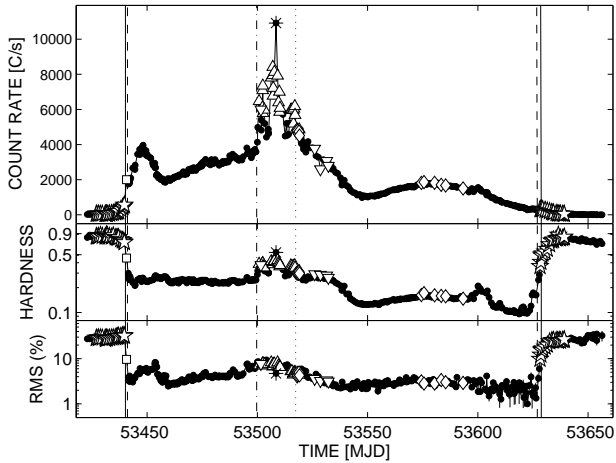}
\caption{Upper panel: lightcurve of the 2005 outburst of GRO J1655-40. Middle panel: evolution of the spectral hardness. Bottom panel: evolution of the integrated fractional rms. The vertical lines mark the state transitions. First solid line: LHS to SIMS. First dashed line: SIMS to HSS. Dot-dashed line: HSS to ULS. Dotted line: ULS to HSS. Second dashed line: HSS to HISM. Second solid line: HIMS to LHS. The symbols are the same as in Fig. \ref{fig:RID}}\label{fig:licu}
\end{center}
\end{figure}
%----------------------------------------------------------------------------------

\subsection{GRO J1655-40}

GRO J1655-40 was discovered with the Compton Gamma Ray Observatory during an outburst that started on July 27, 1994 (\citealt{Zhang1994a}). Radio observations revealed superluminal jets that allowed a distance determination of 3.2 $\pm$ 0.2 kpc (\citealt{Hjellming1995}). Periodic dips have been observed in the optical and X-ray light curves during outburst (\citealt{Bailyn1995}, \citealt{Kuulkers1998}), and optical photometry and spectroscopy in quiescence provided accurate measurements of the inclination of the system (69.5$^\circ \pm$ 0.1$^\circ$, \citealt{Orosz1997}) and the mass of the black hole (6.3 $\pm$ 0.5 M$\odot$; \citealt{Greene2001}). 
GRO J1655--40 underwent two other major outbursts, one started in April 1996 (\citealt{Remillard1996})  and the other started in February 2005 (\citealt{Markwardt2005}). 

%Da Beike-----
From variability studies of RXTE data spanning the 1996--1997 outburst, \cite{Remillard1999} found four types of QPOs between 0.1 Hz and 300 Hz. Three of these QPOs had relatively stable central frequencies, whereas the central frequency of the fourth QPO varied over the range 14--28 Hz. The high-frequency QPO at 300 Hz was later confirmed by \cite{Strohmayer2001}, who also detected a QPO at 450 Hz (the highest frequency QPO seen to date from a black hole, see \citealt{Belloni2011}). Using the same data as \cite{Remillard1999}, \cite{Sobczak2000} examined correlations between the properties of the 14--28 Hz QPO and the spectral parameters, finding that these QPOs were only detected when the hard component in the X-ray spectrum contributed more than 20\% to the 2--20 keV flux, and that the QPO frequency generally increased as the disc flux increased.
 %-------------

In 2005 RXTE monitored a new outburst of the source from a very early stage, observing it on daily basis and covering  the full outburst phase.  
%da Beike----
 \cite{Saito2006} reported on the spectral evolution along the entire outburst, while \cite{Shaposhnikov2007a} presented a study of the spectral and timing evolution during the early stages of the 2005 outburst. \cite{Debnath2008} performed a timing analysis to study the evolution of the variability as a function of the spectral hardening. \cite{Chakrabarti2008} studied LFQPO in GRO J1655-40 observed during the outburst rise and decay in order to probe the origin of the frequency variations of the oscillations.
 
Similar to the 1996 outburst, the 2005 outburst displayed a bright \textit{ultra-luminous state} (ULS, also called \textit{anomalous state}, see \citealt{Belloni2010}, that in the case of GRO J1655-40 roughly corrensponds to the \textit{steep power-law state}, see \citealt{Remillard1996}); it lasted $\sim$ 20 days. % Hiemstra et al. (in prep.) studied in detail the evolution of the timing properties of the source during this state in the 2005 outburst.
%%-----

\section{Observations and data analysis}\label{sec:observations} 

We examined RXTE/PCA public archival observations of GRO J1655-40 obtained during the 2005 outburst. For our timing analysis we only selected observations for which a narrow feature and/or a significant broad peaked component could be detected in the PDS. Our final sample includes a total of 92 observations. 

The PCA data we discuss here were obtained from 2005 February through November and are now part of the public RXTE archive. The PCA data modes employed for most of these observations included a high-time-resolution event mode recording events above PCA $\sim$15 keV (channel 36) and a single-bit mode covering the lower energies from $\sim$2 to $\sim$15 keV (channels 0-35). 

\begin{figure}
\begin{center} 
\includegraphics[width=9.3cm]{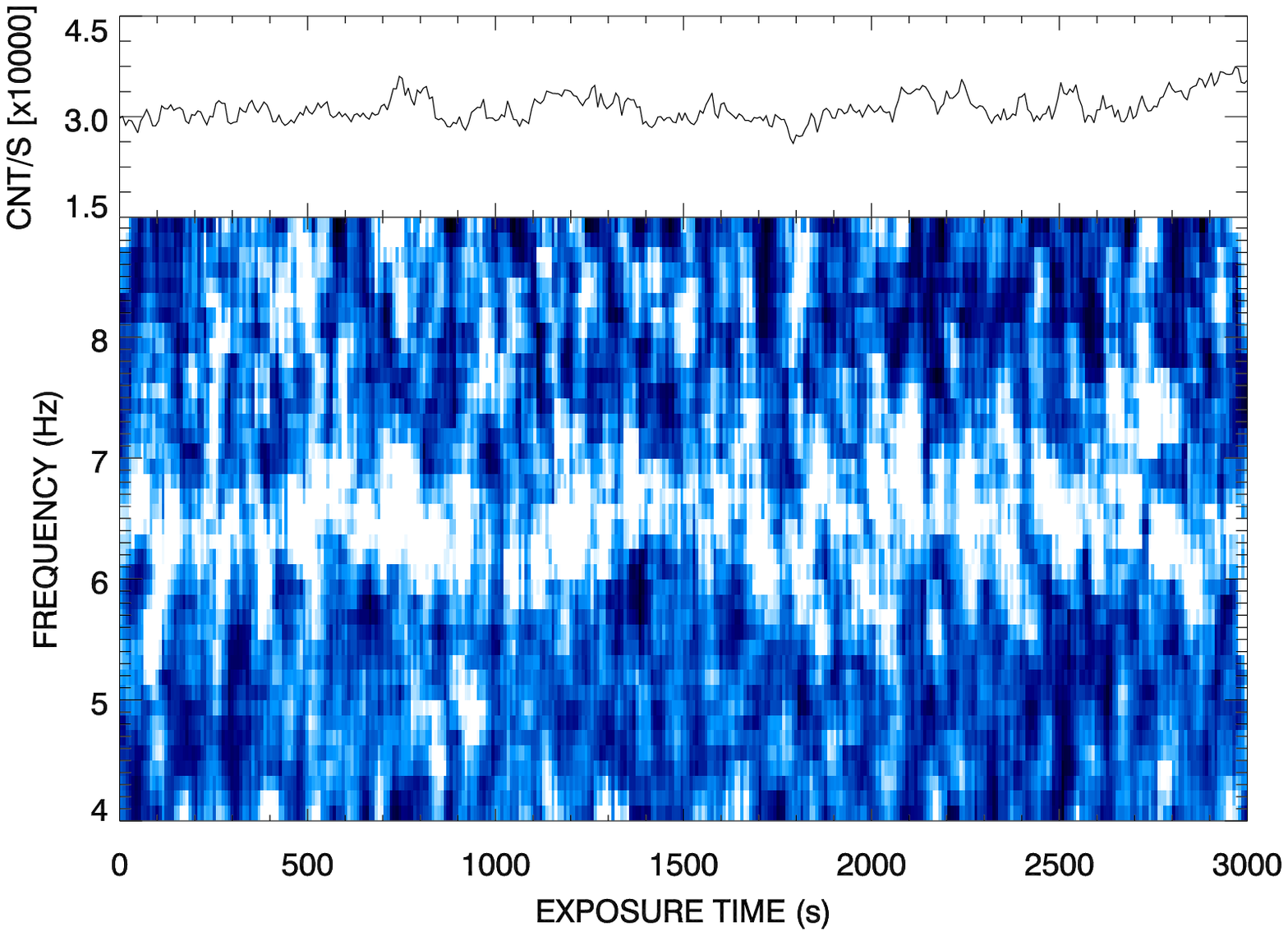}
\caption{Spectrogram for observation \#42 with a time resolution of 8s. The poissonian noise has been  subtracted and the power is normalized according to Leahy et al. (1983). Only the first section of the observation is shown.}\label{fig:DPDS}
\end{center}
\end{figure}

As part of our analysis for each observation  we computed power spectra using custom software under \textsc{IDL} in the energy band 2-26 keV (channel 0 to 62)\footnote{This choice is intended to maximize the S/N of the QPO by including only the energy band where LFQPOs are significant. See \cite{Casella2005} and \cite{Rodriguez2004} and \cite{Rodriguez2008}}. We used 64s intervals and a Nyquist frequency of 1024 Hz. We then averaged the individual spectra for each observation. 
We did not subtract the contribution due to Poissonian noise before fitting the PDS as we experienced difficulties in applying the standard correction formula proposed by \cite{Zhang1995} and \cite{Jahoda2006}. In most of the observations taken during the ULS the source luminosity is high and this results in a lower Poissonian noise level\footnote{The Poissonian noise level depends on the count rate and becomes lower as the source flux increases. The Poissonian noise has been modeled for RXTE by \citealt{Zhang1995}.}. The PDS were normalized according to \cite{Leahy1983} and converted to square fractional rms (\citealt{Belloni1990}). The integrated fractional rms\footnote{We define the integrated fractional rms as the rms integrated over a certain frequency band.} was calculated over the 0.1--64 Hz frequency band. We produced a hardness-rms diagram (see Fig. \ref{fig:RID}, middle panel) following \cite{Belloni2010} and using the method detailed in \cite{Munoz-Darias2011}, we produced an rms-intensity diagram (RID, \ref{fig:RID}, bottom panel). We also examined the PDS in the form of a spectrogram (see Fig. \ref{fig:DPDS}), computing fast Fourier transforms of 8s-long windows of data. In some cases, we used shorter time windows to better follow the evolution of a narrow feature and to explore the evolution of the overall PDS shape over time.

PDS fitting was carried out with the standard {\sc xspec} fitting package by using a one-to-one energy-frequency conversion and a unit response. Following \cite{Belloni2002}, we fitted the noise components with a number of broad Lorentzian shapes (1 to 4 components), one usually zero-centered and the remaining ones centered at a few Hz. The QPOs were fitted with a variable number of Lorentzians depending on the presence of harmonic peaks. A constant component was added to all the PDS to take into account the contribution of the Poissonian noise. 

{\sc Standard 2} mode data, with a 16s-time resolution and suitable for the spectral analysis, were used to create background and dead-time corrected spectra.
We extracted energy spectra for each observation using the standard RXTE software within \textsc{heasoft V. 6.12}. Only data coming from Proportional Counter Unit 2 (PCU2) of the PCA were used for the analysis, as it was the only unit that was active during all observations. A systematic error of $0.6\%$ was added to the PCU2 spectra to account for calibration uncertainties\footnote{See http://www.universe.nasa.gov/xrays/programs/rxte/pca/doc/rmf/pcarmf-11.7 for a detailed discussion on the PCA calibration issues.}. 
We accumulated background corrected PCU2 rates in the {\sc Standard 2} channel bands A = 4 - 44 (3.3 - 20.2 keV), B = 4 - 10 ( 3.3 - 6.1 keV) and C = 11 - 20 (6.1 - 10.2 keV) to produce the hardness-intensity diagram (HID) shown in Fig. \ref{fig:RID} (upper panel). The hardness is defined as H = C/B (\citealt{Homan2005a}). 

%----------------------------------------------------------------------------------
\begin{figure*}
\begin{center}
\includegraphics[width=17.5cm]{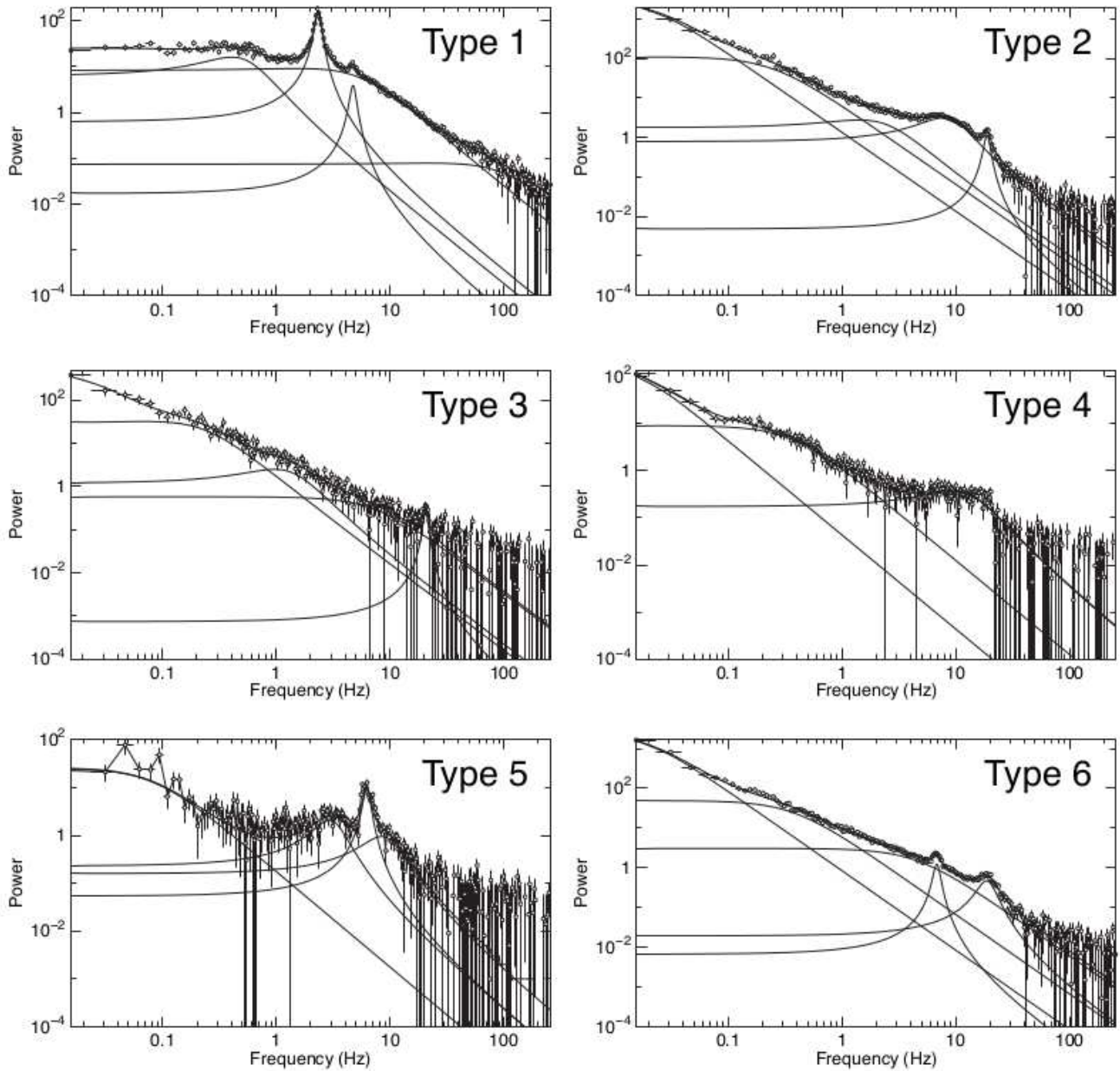}
\caption{Selection of PDS observed during the 2005 outburst of GRO J1655-40. From PDS type 1 to 6 we show Obs. \#27, \#36, \#55, \#65, \#28, \#42 respectively.  PDS Type-6 (Obs. \#42) shows both a type-B QPO at $\sim$6Hz and a type-C QPO at $\sim$18 Hz. The solid curves represent the best fit to each PDS, and its individual components. Notice that  the PDS are normalized according to  Leahy et al. (1983). 
}\label{fig:PDS}
\end{center}
\end{figure*}
%----------------------------------------------------------------------------------

%----------------------------------------------------------------------------------
%----------------------------------------------------------------------------------

\section{Results}\label{sec:results}

In this section we first describe the general behaviour of GRO J1655-40 during the full outburst.  Then we focus on the timing analysis of the observations showing QPOs and/or broad peaked components. %We will study the evolution of the PDS during the outburst in order to clarify the different nature of the two simultaneous QPOs observed in the ULS. 

\subsection{Outburst evolution and the ultra-luminous state }

The HID  in Fig. \ref{fig:RID} (upper panel) shows a rather typical outburst evolution, with the addition of a complex ULS (the upper diagonal extension, above ~2000 cts/s). For a comparison with HIDs of other BHXBs we refer to \citealt{Dunn2010} and \cite{Fender2009}. The ULS is also clearly visible in the RID (the long extention between 5 and 10\% rms levels crossing the diagram from the left to the upper right corner of the plot). The spectral classification of the observations in our sample was performed following the criteria outlined in \cite{Belloni2011}. In addition to the ULS, we identified four different spectral states: low-hard state (LHS), hard-intermediate state (HIMS), soft-intermediate state (SIMS), and high-soft state (HSS).

For a period of about a month after the beginning of the RXTE daily monitoring, GRO J1655--40 remained in the LHS (see Fig. \ref{fig:licu}). After this the source made an extremely fast transition (in about one day) to the SIMS. Even though it was expected, the HIMS, was not sampled by RXTE during this transition, while the subsequent SIMS lasted only one day. \cite{Joinet2008} reported a the presence of a short HIMS (but no SIMS) using data from the INTEGRAL satellite.  After the SIMS GRO J1655--40 reached the HSS, where it remained for about two months, with minor spectral and luminosity changes. This was followed by a period of $\sim$20 days, during which the source was in the ULS. Intense flaring on time scales of hours to days and significant spectral changes were observed during this period (see Fig. \ref{fig:licu}). The count rate also reached its maximum in the ULS ($\sim$11150 counts/s/PCU in the 2-26 keV energy range). After the ULS the source made a transition to the HSS; the count rate decreased continuously for $\sim$ 20 days, while the hardness ratio remained almost constant. The duration of the HSS was four months. Differently from what happened during the softening phase (upper horizontal branch in the HID, \ref{fig:RID}, upper panel), RXTE was able to catch the short-lived (few days) HIMS during the hardening phase (bottom horizontal branch in the HID). Finally, the source moved to the LHS (arrow in the inset of  Fig. \ref{fig:RID}, bottom panel) and reached quiescence after about two months.

The exact position of the transitions from and to LHS can be easily identified through the RID. They form sharp breaks between the branches traced out during  the HIMS and the LHS (see Fig. \ref{fig:RID}, bottom panel). In the HID these transitions appear smoother.

\subsubsection{Evolution of the timing properties}\label{sec:timing}

%Our timing analysis reveals both common and uncommon features during the outburst. %In particular, peculiar PDS appear in the ULS, including a unique PDS showing two non-harmonically related QPO peaks. 
To study the fast time-variability properties of GRO J1655-40, we divided the PDS in our sample (i.e. only those showing QPOs or peaked components) into different groups based on the type A/B/C classification (\cite{Wijnands1999}, \cite{Casella2005} and \cite{Motta2011a}). All the narrow features could be classified as one of the three types of QPOs, even though in a few cases the classification remains ambiguous. Our PDS and QPO classification is detailed in Tab. \ref{tab:observations}.

Among  the 92 observations we analyzed, we identify 6 different types of PDS. In Fig. \ref{fig:PDS} we show an example for each type. The reader should notice that the following types of PDS do not exactly correspond to the ones described in \cite{Belloni2010}, even though some  correspondences can be found (see below).
\begin{itemize}

\item {\bf Type 1:} the PDS shows two main components: a strong flat-topped noise - that can be fitted by three broad Lorentzians - and one or more QPO peaks. When more than one QPO peak is observed, they are harmonically related. The fundamental peak moves in the 0.1--13 Hz frequency range (0.1--3 Hz during the rise phase of the outburst and 0.2--13 Hz during the decay) and is strong and narrow (\textit{quality factor}\footnote{The \textit{quality factor Q} is defined as the ratio of the frequency of a Lorentzian and its FWHM. Conventionally, a QPO is defined as a feature with a quality factor larger than 2.} Q $>$ 4). This type of PDS corresponds to PDS 1 and 2 in \citealt{Belloni2010}, Sec. 3.3.1.
The integrated fractional rms of the PDS is always higher than 10\%. These PDS are observed in the LHS during the outburst rise and both during the HIMS and LHS in the decay. %This is a typical HIMS PDS.%According to the ABC classification we identified these peaks as type-C QPOs. 

\item {\bf Type 2:} the PDS shows a broad noise component that can be described by three broad Lorentzians, a narrow QPO peak at $\sim$ 20 Hz and a peaked noise component with a characteristic frequency between 6 and 8 Hz. The broad peaked component appears in the PDS in correspondence to local peaks in the count rate. The lightcurve shows significant flaring also on short timescales (hundreds of seconds). These PDS show typical fractional rms in the range 4-10\% and they are observed during the ULS. %The inspection of the spectrogram shows that the rms of both peaked noise component and type-C QPO correlate with the count rate. We tentatively classify the peaked noise component to a type-B QPOs-like feature.  %The high quality factor and the frequency range that they cover allow to classify the narrow peaks as type-C QPOs. 

\item {\bf Type 3:} a weak (but significantly detected) and narrow QPO appears at frequencies between 20 and 28 Hz and is associated to a weak flat-topped noise component with a break at few Hz plus a stronger power-law noise component dominating at lower frequencies. These PDS are found in the HSS, some close to the transition between ULS and HSS and others in a more typical (softer) HSS. The QPOs have lower frequencies (19-21 Hz) and appear weak and broad (Q $\sim$ 3 - 10) near the ULS/HSS transitions, while they show higher frequencies (25 - 28 Hz) and are strong and narrow (Q $\sim$ 15 -39) in softer parts of the HSS. Averaging the PDS of the two sub-groups we find a peak at 20.1 Hz for the first group and 27.2 Hz for the second (soft HSS) group, with Q = 4.4 and Q = 16.6 respectively. In both cases the break associated to the noise appears sharper in the averaged PDS than in the individual ones.  This PDS corresponds to PDS 5 in \cite{Belloni2010}.%It is worthy noticing that type-C QPOs are usually observed only in the hard states, while in the HSS only type-A QPOs are found. %We classified those QPO as type-C QPOs.

\item {\bf Type 4:} the PDS show a peaked noise component (slightly weaker and broader with respect to the PDS of type 2)  with no significant QPO. Differently from PDS type 2, the type 4 PDS do not appear associated to local maxima in the light curve, even though the lightcurve shows moderate flaring on short timescales (hundreds of seconds). These PDS are mostly found in the HSS close to the transition from ULS and HSS. One case (Obs. \#30) is observed in the ULS.

\item {\bf Type 5:} this type of PDS is found within Obs. \#28, which  consists of two RXTE orbits. The inspection of the spectrogram from the entire observation revealed that the shape of the PDS is variable during the observation. During the first orbit (first section of the observation) the  light curve shows low variability and the PDS reveals a strong flat-topped noise component and no significant QPO. In the first half of the second orbit (second section, $\sim$4900 to $\sim$5900 second after the observation start time) the  PDS show a weaker power-law noise component and two harmonically related peaks, at $\sim$3 and $\sim$6 Hz. In the second half of the second orbit (third section), the PDS shows power-law noise and a strong peak at $\sim$6 Hz with a faint sub-harmonic at $\sim$3 Hz. 
The first section of Obs. \#28 results in a PDS shape very similar to type 1 PDS (even though no QPO is detected), while the second and third section of Obs. \#28 result in the PDS shape that we classify as type 5 (see Fig. \ref{fig:PDS}).
During the entire second  orbit, the light curve shows dips\footnote{We studied the properties of those dips and we verified the absence of hardening, therefore we exclude an absorption origin (see \citealt{Kuulkers1998}).} 4-8 s long, during which the  PDS takes the form of strong flat top noise with no QPO.  In Fig. \ref{fig:PDS} we only show the PDS extracted from the second part of the second orbit, where the QPO centroid peak is stronger. This PDS is observed in the SIMS and corresponds to PDS 3 in \cite{Belloni2010}.%This \textit{hybrid} behavior is reflected by the RID, as this observation lays just on the borderline between HIMS and SIMS, as the total rms equals 10\%. This has been already observed in GX 339-4 (see \citealt{Munoz-Darias2011}) . 

\item {\bf Type 6:}  our sample contains a very peculiar PDS (Obs. \#42) that shows two non-harmonically related peaks at 6.84 $\pm$0.03 Hz and 18.7$\pm$0.1 Hz superimposed on strong power-law noise. These two simultaneous non-harmonically related peaks were first reported by \cite{Homan2005c}.  A careful inspection of the spectrogram confirms that the two peaks are both always present during the full length of the observation. 
The peak at $\sim$ 6 Hz has  fractional rms  = 0.81\% and is narrow (Q = 6.8) and moves in frequency on short timescales (tens of seconds, see Fig. \ref{fig:DPDS}). The peak at higher frequency is broader (Q = 2.4) and has fractional rms =  1.4\%. This peak is weaker than all the other peaks observed at similar frequencies during the 2005 outburst. The spectrogram shows that the width of this peak is intrinsic (i.e. it is not due to fast frequency variations along the observation\footnote{On a time scale that is equal or longer than 8s, i.e. the lenght of the lightcurve stretches used to produce the spectrogram}). As shown in Fig. \ref{fig:RID} (middle and bottom panel), the integrated fractional rms of the whole PDS is significantly lower then the previous and following observations and is close to 5\%. The light curve of this observation shows strong flaring on short time scales.  This observation belongs to the ULS and correspond to the highest count-rate observed during this outburst.
%The ABC scheme allows to classify the peak at lower frequency as a type-B QPO. The peak at higher frequency is broader and fainter with respect to all the peaks observed at $\sim$ 20 Hz during the outburst (classified as type-C QPOs). We tentatively classify this peak as a type-C QPO. 
\end{itemize}

\subsubsection{Relations between rms and frequency}

%----------------------------------------------------------------------------------
\begin{figure}
\begin{center}
\includegraphics[width=9.0cm]{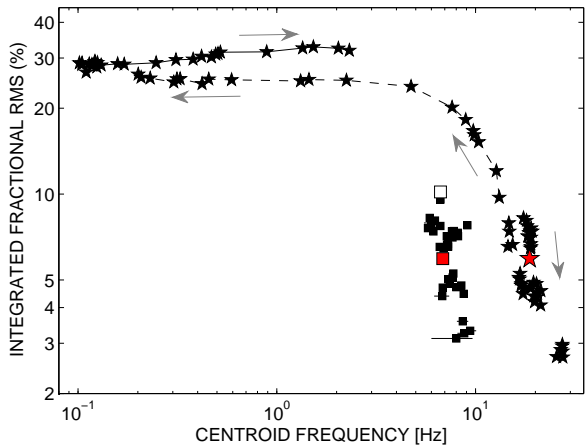}
\caption{QPO centroid frequency versus integrated (0.1--64 Hz) fractional rms. Each point corresponds to a single RXTE observation. The grey arrows mark the QPO frequency evolution. The solid line joins as a function of time the observations showing QPOs in the LHS  during the rise of the outburst; the dashed line joins the observations showing QPOs observed in the LHS and the HIMS during the decay of the outburst. 
We included in the plot both the observations showing a QPO or a peaked noise component. Stars are type C QPOs and squares type B QPOs. The red star corresponds to the type-C QPO observed simultaneously with the type-B QPO (red square) in PDS type 6 (Obs. \#42). The white square corresponds to the the type-B QPO observed in PDS type 5 (Obs. \#28). The black squares correspond to the peaked noise components observed in the ULS (type 2 PDS).}\label{fig:rms_fr}
\end{center}
\end{figure}
%----------------------------------------------------------------------------------

Following \cite{Casella2005} and \cite{Motta2011a}, we plot in Figure \ref{fig:rms_fr} the integrated fractional rms of each PDS versus the centroid frequency of the QPO. This is a useful method for differentiating between different types of PDS as the integrated fractional rms is known to correlate well with the frequency of some type of LFQPOs  (see e.g. \citealt{Casella2005}, \citealt{Motta2011a}, \citealt{Munoz-Darias2011a}, \citealt{Kalamkar2011}).  In particular, we want to investigate whether the two simultaneous and non-harmonically related peaks detected at the peak of the outburst (type-6 PDS in \ref{fig:PDS}) can be identified as type A, B, or C QPOs.
For completeness, we also include the points corresponding to peaked noise components observed in PDS type 2 and 4 in  Figure \ref{fig:rms_fr}.

Figure \ref{fig:rms_fr} shows that two groups of points can be identified and associated to two different types of QPOs. The QPOs observed in the LHS and HIMS during the rise and decay of the outburst are given as function of time, joined by the solid and dashed line, respectively.

\begin{itemize}
\item A large group of points (black stars) forms a curved track that covers a large frequency ($\sim$0.1-30 Hz) and rms range (3-30\%).  These points correspond to all the narrow peaks observed  in PDS type 1, 2 and 3. The points coming from the rise of the outburst (i.e. along the upper horizontal branch of the HID) are found at higher integrated fractional rms and at a frequency below $\sim$3Hz. At frequencies above 10 Hz we observe two parallel tracks corresponding to the rise (higher frequency) and decay phase of the ULS\footnote{We call \textit{rise} of the ULS all the observations of the ULS before Obs. \#42 (outburst peak) and \textit{decay} all the ULS observations after Obs. \#48.}. As Fig. \ref{fig:licu} shows, the integrated fractional rms is generally higher during the rise than during the decay of the ULS.
According to the ABC classification (\citealt{Casella2005}), the properties of these peaks, summarized in Tab. \ref{tab:timing_sum}, allow us to classify them as type-C QPOs. 

\item Another group of points (black squares) corresponds to the peaked noise component observed in PDS type 2 and 4. They form a different track along which the frequency weakly anti-correlates with the integrated fractional rms. They are mostly found in the 4-10\% integrated fractional rms range, with few points at $\sim$ 3\% rms. This track includes the QPO observed in PDS type 5 during the SIMS (Obs. \#28), marked by a white square in Fig. \ref{fig:rms_fr}. This last peak can be classified as a type-B QPO according to the ABC scheme\footnote{Following \cite{Casella2004}, the peak takes the form of a type-B \textit{cathedral} in the second section of Obs. \#28 (see description of type 5 PDS in  Sec. \ref{sec:timing}) and of a standard type-B in the third section of Obs. \#28.}. 
Even though the peaked noise components cannot be properly defined QPOs given their broadness, their properties (see Tab. \ref{tab:timing_sum}) suggest similarities to type-B QPOs (see also \S \ref{sec:discussion}) . 

\item The simultaneous QPOs detected at the peak of the outburst at $\sim$18Hz and $\sim$7 Hz (type-6 PDS, Obs. \#42, red star and red square in Fig. \ref{fig:rms_fr}) fall  on the first and second group of points, respectively. The properties of these two peaks allow us to classify them as a type-C QPO and a type-B QPO and shows that a type-C and a type-B QPO can be detected simultaneously. The fast frequency variations shown by the peak at lower frequencies (see Fig. \ref{fig:DPDS}) further suggest its type-B nature \citep{Nespoli2003}.
\end{itemize}

%%_____________________BEGIN________TABLE 1____________________________%%
\begin{table*} 
\begin{center} 
\caption{Summary of the timing properties of GRO J1655-40 during the 2005 outburst.  The integrated rms is measured between 0.1 and 64 Hz.}\label{tab:timing_sum} 
\begin{tabular}{c c c c c c } 
\hline
	&	Type C	&	Type C	&	Type C	&	Type-B	&	Peaked component	\\
	&	LHS/HIMS	&	ULS	&	HSS	&		&		\\
\hline											
\hline											
Frequency (Hz)	&	0.1 - 13	&	14 - 20	&	20 - 28	&	$\sim$ 7	&	6 - 9	\\
Q	&	3 - 9 	&	3 - 11	&	3 - 39	&	3 - 7 	&	0.4 - 1.4	\\
integrated rms (\%)	&	10 - 30	&	4 - 8	&	2- 5 	&	5 - 10	&	5 - 10	\\
noise	&	strong flat top	&	weak flat top + red	&	weak flat top + red	&	red 	&	red 	\\
\hline
\end{tabular}
\end{center} 
\end{table*} 
%%_____________________END__________TABLE_1____________________________%%

\section{Discussion}\label{sec:discussion}

We have analyzed 507 RXTE observations of the BHXB GRO J1655-40 from its 2005 outburst. We selected all the observations  showing LFQPOs in order to investigate their properties and evolution throughout the outburst of the source. This sample consisted of a total of 92 observations.

%{\bf I suggest removing these this paragraph/sentence - it doesn't add much to the discussion and some of the statements are quite debatable} Despite the complex outburst of GRO J1655-40, the inspection of the HID and the RID allows us to  separate the ULS from the rest of the outburst thanks to a detailed spectral classification (see \citealt{Belloni2011}) and to put this source in the context of the canonical states scenario. 

%In this particular case this operation is made difficult by the complexity of the ULS, which extends by a factor of two the flux range usually spanned during BHT's outbursts. %However, fast-time variability well tracks different accretion regimes (see \citealt{Munoz-Darias2011a} and allows to recognize the canonical spectral states without the intervention of any spectral information. 
%The use of both diagrams is particularly useful in cases - as the one presented here -  where the HID is of difficult interpretation (see also Soleri et al. 2012, submitted). 

All LFQPOs could be classified using the ABC scheme and the rms-frequency diagram. This includes those LFQPOs detected in the peculiar PDS types observed in the ULS. %Most of the features observed in the ULS enter in the same classification . The use of the rms-frequency diagram allowed us to confirm the simultaneous detection of a type B and type C QPO. 
Most of the LFQPOs are type-C QPOs.  54\% of these are observed during the LHS and the HIMS, showing the same frequency evolution and properties already observed in many other sources (see e.g. GX 339-4,  \citealt{Motta2011a}; XTE J1859+226, \citealt{Casella2004}; MAXI J1659-152 \citealt{Kalamkar2011}). These QPOs have frequencies between 0.1 and 13 Hz (0.1 to 3 Hz during the rise phase of the outburst and 0.2--13 Hz during the decay phase). About 30\% of the type-C QPOs are found in the ULS at frequencies ranging from $\sim$14 Hz and $\sim$20 Hz and they usually appear together with a peaked noise component with characteristic frequency of $\sim$7 Hz. Interestingly, 16\% of the type-C QPOs are observed in the HSS. Type-C QPOs are typically only observed in spectrally harder states, although similar QPOs have also been found in the HSS states of XTE J1550--564 \citep{Homan2001} and H1743--332 \citep{Homan2005b}.  Of the type-C QPOs in the HSS, some are observed very close to the transition from the ULS to the HSS and their frequency is about 20 Hz. Other type-C QPOs are found far from the transition and their frequency ranges from 25 Hz to 28 Hz. These QPOs, already reported by \cite{Remillard1999} are among the highest  frequency type-C QPOs observed in  BHXBs (see \citealt{Vignarca2003} for the case of XTE J1748-288). The relations traced out in the frequency-rms diagram by the type-C QPOs observed in the ULS and HSS appear to be continuations of the branch traced out by the HIMS type-C QPOs, although they are shifted toward slightly higher frequencies.

Only one LFQPO displayed all the standard properties of a type-B QPO (i.e. the single SIMS observation during the outburst rise). However, in several observations of the ULS we detected a peaked noise component that shares some of the characteristics of type-B QPOs (although they remain different features, mainly because of their width). This noise component had a characteristic frequency  between 6 Hz and 8 Hz and they appear to be related to significant increases in the count rate, similarly to what was observed by e.g. \cite{Motta2011a} in the case of type-B QPOs. In the frequency-rms diagram the noise components trace out a track that is well separated  from type-C QPOs and that also includes the type-B QPO detected in the SIMS. This track also corresponds to the region where type-B QPOs are usually observed in other sources (see e.g.  \citealt{Casella2004}). 

\begin{figure}
\begin{center}
\includegraphics[width=8.5cm]{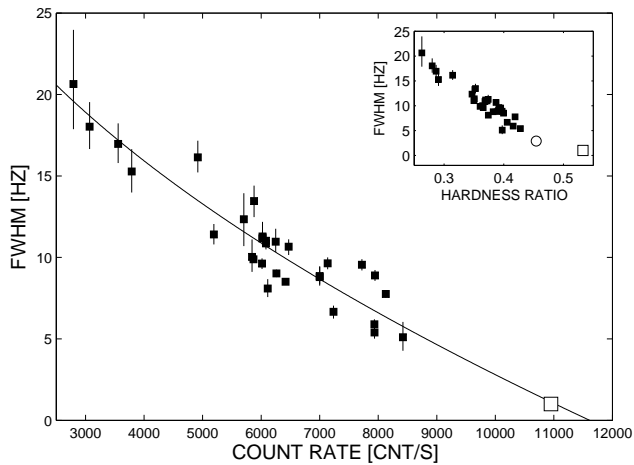}
\caption{Relation between the FWHM of the peaked noise component in the  type 2 PDS and the count rate. The white square represents the type-B QPO observed at the peak of the outburst, simultaneously with the type-C QPO. The relation is well described by a simple polynomial of the form $y = Ax^B + C$, where $A= -0.28$, $B = 0.52$ and $C=37.29$. The inset shows the FWHM of the noise component as a function of the hardness ratio, where the white dot represents the type-B QPO observed in the SIMS ( Obs.  \#28.)  }\label{fig:width}
\end{center}
\end{figure}

Following the evolution of the peaked noise component along the ULS, it appears that this broad component evolves into a narrow type-B QPO as the count rate rises. Figure \ref{fig:width} shows  the dependence of the peaked noise FWHM on count rate. In the inset of Fig. \ref{fig:width} we show the relation between the FWHM of the peaked noise as a function of the hardness ratio and  we also include the type-B QPO observed at the peak of the outburst (white square). The width of the peaked noise component anti-correlates well with the count rate.  %{\bf I'm not sure how this connects to Figure 6} 
As already suggested by \cite{Motta2011a}, the count rate could rise as a consequence of an increase in the local accretion rate, sudden changes in the geometry or radiative efficiency, or even the appearance of an  additional component contributing to the (hard) emission (i.e. jet from the inner regions of the system). Our result might indicate that, at least for what concerns the ULS, if the increase in count rate is strong enough (in other words if the variation in count rate - not the absolute count rate itself  - is large enough) the mechanism responsible of the production of type-B QPOs is triggered and the PDS will show either a peaked noise component with characteristic frequency $\sim$ 6-8 Hz or a proper type-B QPO (as in observation \#28, see below), depending on the count-rate level that is reached. At lower count rates the peaked noise component may simply become too broad (and weak) to be detectable. This result shows that under certain conditions sudden count rate increases might constitute a key ingredient in the appearance of  type-B QPOs (see also \citealt{Stiele2012} for a discussion on the topic).

The timing properties of the ULS, in particular the presence of these type-B like features in the PDS, suggest that most of the ULS  can be regarded as a sort of SIMS covering a higher flux range. We note that in the RID the ULS falls in the region where the SIMS is usually observed (see \citealt{Munoz-Darias2011} and \citealt{Motta2011a}, \citealt{Belloni2011}). However, \cite{Belloni2011} defines the SIMS on the basis of the presence of a type-B QPO. During the ULS of GRO J1655-40 both type-C QPOs and type-B QPOs (or type-B like features) appear, as well as a strong red noise. Therefore, even though the ULS shares the characteristics of the SIMS, it still remains significantly different from it. 

The unique PDS observed at the peak of the outburst shows two simultaneous non-harmonically related peaks at $\sim$7 Hz and $\sim$ 18 Hz that we could classify as a type-B and type-C QPO, respectively. The narrow peak at 7 Hz falls on the same relations traced by the peaked noise components in a rms-frequency diagram, while the broader peak at 18 Hz lies on the track formed by the type-C QPOs observed all over the outburst. The peak at 7 Hz also shows significant frequency variations between  6 and 8 Hz on a short time scale (few seconds) that are typical of type-B QPOs (see \citealt{Nespoli2003}). 
It is important to stress that this PDS is observed in the ULS, when the count rate reached its  maximum during the 2005 outburst. Reaching a very high cound rate (larger than $\sim$10000 cnt/s/PCU in the 3-20 keV energy band) and the source being in the ULS could be necessary conditions for the peak peak at $\sim$7 Hz to become a narrow feature (Q $\leq$ 2) and therefore for the appearance of two simultaneous different types of QPOs.

%----------------------------------------------------------------------------------
%----------------------------------------------------------------------------------

A few additional remarks on the nature of the two simultaneous type-B and type-C QPOs observed at the outburst peak need to be made.

\begin{itemize}
\item The low-frequency noise associated to the two simultaneous peaks is of a hybrid kind: power-law noise plus a weak flat-topped noise superimposed on it. The red noise is typically associated to type-A and -B QPOs, while flat-topped noise is usually observed together with type-C QPOs.

\item The QPO classified as type-C is broader than the other type-C QPOs observed during the outburst (see Tab. \ref{tab:observations}, e.g. observations \#41 and \#42) and it appears similar to a type-A QPO in shape (see \citealt{Casella2004}, \citealt{Casella2005} and \citealt{Motta2011a}). Nevertheless, its properties indicate a type-C nature. In addition, we note that  type-C QPOs sometimes become broad and quite weak close to the transition between HIMS and SIMS (see Homan et al. in prep.)

\item The integrated fractional rms of this PDS is $\sim$5\%, which is significantly lower than the usual values seen for type-C QPOs (i.e. $>$ 5\%), but consistent with the typical rms level at which type-A QPOs are observed (\citealt{Motta2011a}). This observation is very close to the rms limit associated to the transition from SIMS to HSS ($\sim$5\% rms, see \citealt{Belloni2011} and \citealt{Munoz-Darias2011}).
\end{itemize}

Type A and C QPOs usually have significantly different properties, in particular for what concerns the broad-band noise that comes with these two types of oscillations; (strong) flat-top noise in the case of type C QPOs and weak power law or weak peaked noise in the case of type A QPOs. \cite{Motta2011a} showed that type-C and -A QPOs share several properties, the most important one  being the fact that their frequencies follow the same relations as a function of the flux of the hard component in the energy spectrum. Therefore, they could share a common origin. In this scenario type-A and -C QPOs would be two different effects of a same physical process. Some of the QPOs in GRO J1655-40 (e.g. the peak at $\sim$ 18 Hz at the outburst peak and a few of the type-C QPOs observed in HSS) show characteristics of both type-C and type-A QPOs. As discussed in Sec. \ref{sec:timing}, the QPO at 18Hz observed simultaneously with the type-B QPO at the outburst peak shows the behavior of a type-C QPO (e.g. it follows the typical type-C frequency evolution), but the characteristics detailed  above (e.g. intrinsic broadness and faintness) suggest that it also shows some of the properties of type-A QPOs. Therefore it might be regarded as a link between type-C and type-A QPOs.
%{\bf I would stop here. The remainder of this paragraph is not very well written and does not contain a lot of useful information.} 
%Up to now no type-A QPO has been observed at frequencies above $\sim$10Hz, even though we note that the number of type-A QPOs detections is generally  small compared to the number of detections of type-B and type-C QPOs in the PDS of BHXBs. 
In the PDS of GRO J1655-40, we did not detect any type-A QPOs (usually observed in the SIMS), but we found several type-C QPOs associated to an `hybrid' noise (formed by a weak flat top noise component superimposed to a strong red noise) in HSS, where it is uncommon to observe any type of QPO. Unfortunately the complete lack of type-A QPOs in the PDS of GRO J1655-40 makes impossible a  detailed comparison between type-A QPOs and HSS type-C QPOs. 

The simultaneous detection of a type-B and a type-C QPO proves that at least two intrinsically different types of LFQPOs exist. 
The fact that two different types of QPOs can be observed at the same time is of great importance to understand the origin of LFQPOs, since it indicates that they are likely produced by different physical processes which, under opportune conditions, can take place simultaneously.% {\bf I would stop here as well. Last sentence is a bit meaningless} This result opens several points over the physical origin of LFQPOs in black hole candidates  and on the actual processes that could explain the presence of those features and their association to specific broadband noise components (i.e. flat-top noise and type-C QPOs; red noise and type-B/A QPOs). 

\cite{Stella1998} originally proposed that the Lense-Thirring precession could be the mechanism producing LFQPOs in accreting compact objets. \cite{Ingram2009} developed this model in the truncated disk framework (see \citealt{Done2007}), where the precession of a hot flow in the inner regions of the accretion plasma is the origin of type-C QPOs. \cite{Motta2011a} suggested that the same mechanism could also explain the presence of type-A QPOs in the emission of BHCs. The results reported here essentially rule out the possibility that type-B QPOs could arise from the same physical phenomenon - regardless of what this phenomenon is - supporting what was already suggested by \cite{Motta2011a}. 

Our results show that the presence of two simultaneous type-C and -B QPOs in the emission from GRO J1655-40 is possibly related to the ULS. Similar cases have been found in the 1995 outburst of GRO J1655-40 as well as in the PDS of the black hole candidate XTE J1550-564 and H1743-322 (Motta et al in prep.). In all these cases simultaneous type-B and type-C QPOs are observed at the highest count rate end of the ULS. However, only GRO J1655-40 showed a sufficiently long and bright ULS associated to a standard outburst evolution, that offered the possibility to study in detail the evolution of the timing properties.
The investigation of the physical circumstances underlying the presence of two simultaneous different types of LFQPOs is beyond the scope of this work. However, a detailed spectral analysis of the ULS in the 2005 outburst of GRO J1655-40 as support of the timing analysis will be presented in a forthcoming work. 

%Further efforts in shaping a consistent scenario supported by a solid theoretical model able to explain the fast timing properties observed in BHCs as well as the spectral changes typical of these systems, are strongly needed. 

\section{Summary and conclusions}								
						
We report on a timing analysis of 507 RXTE observations of the BHXB GRO J1655--40 during its 2005 outburst. 
We selected all the observations in which a LFQPO was observed and we classified the QPOs according to the ABC scheme. 

Our sample included a unique observation whose PDS shows two simultaneous, non-harmonically related QPOs which we identified as a type-B and a type-C QPO. 
This result proves that in the emission of GRO J1655-40 at least two of the three known types of LFQPOs are intrinsically different phenomena and likely produced by different physical processes. This could be valid by extention for other sources as well. Our results also show that the requirements for the appearance of a type-B or a type-C QPO are usually mutually exclusive, but under opportune conditions, these two different types of QPO can be observed simultaneously.
Of the two simultaneous QPOs, the one classified as a type-C shows some resemblance to a type-A QPOs as well. This weakens the distinction between type-C and -A QPOs, supporting the hypothesis that these two types of LFQPOs may share a similar physical origin.

We also investigated the nature of the broad peaked noise component observed in most PDS of the ULS observations and we found that it is linked to the type-B QPO observed simultaneously with the type-C QPO at the brightest observation of the outburst. This peaked component is present during local maxima in the light curve. Its width correlates well with the count rate, to the extent that it evolves into a narrow type-B QPO when the count rate is higher than a certain value.

\vspace{1cm}
\noindent  The authors would like to thank the anonymous referee for useful comments that contributed to improve the paper.
SM and TB acknowledge support from grant ASI-INAF I/009/10/. The research leading to these results has received funding from the European Community's Seventh Framework Programme (FP7/2007-2013) under grant agreement number ITN 215212 \textquotedblleft Black Hole Universe\textquotedblright. TMD acknowledges funding via an EU Marie Curie Intra-European Fellowship under contract no. 2011-301355. TMB acknowldges support from the Leverhulme Trust.
This research has made use of data obtained from the High Energy Astrophysics Science Archive Research Center (HEASARC), provided by NASA's Goddard Space Flight Center. 

\newpage					
	  																																															
\onecolumn																																																		
\begin{center} 																																																		
\begin{landscape}																																																		
\begin{longtable}{|c|c|c|c|c|c|c|c|c|c|c|c|c|c|} 																																																		
\caption{Power spectral classification and variability parameters. Only observations with evidence of LFQPOs or a peaked noise componet are listed. The PDS are classified according to the creteria detailed in the text (see Sec. \ref{sec:timing})}\label{tab:observations} \\ 																																																		
\endfirsthead																																																		
																																																								
\multicolumn{14}{c}%																																																											
{{\tablename\ \thetable{} -- continued from previous page}} \\ \hline																																																											
\multicolumn{8}{|c|}{ } & \multicolumn{2}{|c|}{Type-C QPOs} & \multicolumn{2}{|c|}{Type-B QPOs} & \multicolumn{2}{|c|}{Peaked noise} \\																																																											
\hline																																																											
\#	&	Time	&	Obs. ID	&	Hardness ratio			&	Count rate			&	rms 				&	State	&	\footnotesize{PDS}	&	Frequency					&	FWHM					&	Frequency					&	Width					&	Frequency					&	FWHM						\\
	&	[MJD]	&		&				&	[cnts/s]			&	[\%]				&		&	\footnotesize{Type}	&	[Hz]					&	[Hz]					&	[Hz]					&	[Hz]					&	[Hz]					&	[Hz]						\\
\hline																																																											
\hline																																																											
\endhead																																																											
																																																											
\hline \multicolumn{14}{c}{{Continued on next page}} \\																																																											
\endfoot																																																											
																																																											
\hline 																																																											
\endlastfoot																																																											
\hline																																																											
																																																											
\hline																																																											
\multicolumn{8}{|c|}{ } & \multicolumn{2}{|c|}{Type-C QPOs} & \multicolumn{2}{|c|}{Type-B QPOs} & \multicolumn{2}{|c|}{Peaked noise} \\																																																											
\hline																																																											
\#	&	Time	&	Obs. ID	&	Hardness ratio			&	Count rate			&	rms 				&	State	&	\footnotesize{PDS}	&	Frequency					&	FWHM					&	Frequency					&	Width					&	Frequency					&	FWHM						\\
	&	[MJD]	&		&				&	[cnts/s]			&	[\%]				&		&	\footnotesize{Type} 	&	[Hz]					&	[Hz]					&	[Hz]					&	[Hz]					&	[Hz]					&	[Hz]						\\
\hline																																																											
\hline																																																											
																																																											
1	&	53427	&	90058-16-05-00	&	0.837	$\pm$	0.007	&	48.3	$\pm$	0.2	&	28.8	$\pm$	0.5		&	LHS	&	1	&	0.105	$_{-	0.003	} ^{+	0.002	}$   &	0.025	$_{-	0.006	} ^{+	0.009	}$   &	 -					&	 -					&	 -					&	 -						\\
2	&	53427.2	&	90428-01-01-01	&	0.826	$\pm$	0.006	&	50.1	$\pm$	0.2	&	28.4	$\pm$	0.4		&	LHS	&	1	&	0.104	$_{-	0.002	} ^{+	0.004	}$   &	0.032	$_{-	0.007	} ^{+	0.006	}$   &	 -					&	 -					&	 -					&	 -						\\
3	&	53427.9	&	90058-16-07-00	&	0.837	$\pm$	0.006	&	54.2	$\pm$	0.2	&	29.1	$\pm$	0.3		&	LHS	&	1	&	0.122	$_{-	0.005	} ^{+	0.002	}$   &	0.035	$_{-	0.007	} ^{+	0.008	}$   &	 -					&	 -					&	 -					&	 -						\\
4	&	53428.1	&	90428-01-01-03	&	0.830	$\pm$	0.006	&	53.9	$\pm$	0.2	&	28.4	$\pm$	0.4		&	LHS	&	1	&	0.117	$_{-	0.003	} ^{+	0.000	}$   &				$<$	0.004	&	 -					&	 -					&	 -					&	 -						\\
5	&	53428.9	&	90428-01-01-04	&	0.859	$\pm$	0.004	&	52.5	$\pm$	0.1	&	29.1	$\pm$	0.2		&	LHS	&	1	&	0.121	$_{-	0.001	} ^{+	0.004	}$   &	0.015	$_{-	0.002	} ^{+	0.004	}$   &	 -					&	 -					&	 -					&	 -						\\
6	&	53429.7	&	90428-01-01-02	&	0.836	$\pm$	0.004	&	55.7	$\pm$	0.1	&	28.9	$\pm$	0.2		&	LHS	&	1	&	0.125	$_{-	0.002	} ^{+	0.002	}$   &	0.019	$_{-	0.002	} ^{+	0.005	}$   &	 -					&	 -					&	 -					&	 -						\\
7	&	53431	&	90428-01-01-05	&	0.831	$\pm$	0.005	&	50.8	$\pm$	0.1	&	28.8	$\pm$	0.3		&	LHS	&	1	&	0.102	$_{-	0.005	} ^{+	0.009	}$   &				$<$	0.002	&	 -					&	 -					&	 -					&	 -						\\
8	&	53431.2	&	90058-16-06-00	&	0.834	$\pm$	0.008	&	51.1	$\pm$	0.2	&	26.7	$\pm$	0.6		&	LHS	&	1	&	0.110	$_{-	0.003	} ^{+	0.003	}$   &	0.028	$_{-	0.008	} ^{+	0.009	}$   &	 -					&	 -					&	 -					&	 -						\\
9	&	53431.6	&	90428-01-01-06	&	0.828	$\pm$	0.008	&	51.2	$\pm$	0.2	&	28.3	$\pm$	0.6		&	LHS	&	1	&	0.119	$_{-	0.002	} ^{+	0.002	}$   &	0.021	$_{-	0.005	} ^{+	0.006	}$   &	 -					&	 -					&	 -					&	 -						\\
10	&	53431.7	&	90428-01-01-07	&	0.866	$\pm$	0.007	&	47.8	$\pm$	0.2	&	28.6	$\pm$	0.4		&	LHS	&	1	&	0.114	$_{-	0.003	} ^{+	0.003	}$   &	0.042	$_{-	0.007	} ^{+	0.009	}$   &	 -					&	 -					&	 -					&	 -						\\
11	&	53431.8	&	90428-01-01-08	&	0.869	$\pm$	0.008	&	49.3	$\pm$	0.2	&	27.9	$\pm$	0.6		&	LHS	&	1	&	0.124	$_{-	0.007	} ^{+	0.006	}$   &	0.023	$_{-	0.005	} ^{+	0.012	}$   &	 -					&	 -					&	 -					&	 -						\\
12	&	53431.9	&	90428-01-01-09	&	0.853	$\pm$	0.004	&	51.6	$\pm$	0.1	&	28.3	$\pm$	0.2		&	LHS	&	1	&	0.131	$_{-	0.003	} ^{+	0.001	}$   &	0.025	$_{-	0.004	} ^{+	0.004	}$   &	 -					&	 -					&	 -					&	 -						\\
13	&	53432.8	&	90428-01-01-10	&	0.834	$\pm$	0.004	&	62.2	$\pm$	0.1	&	28.6	$\pm$	0.2		&	LHS	&	1	&	0.158	$_{-	0.002	} ^{+	0.002	}$   &	0.030	$_{-	0.004	} ^{+	0.005	}$   &	 -					&	 -					&	 -					&	 -						\\
14	&	53433	&	91404-01-01-00	&	0.831	$\pm$	0.005	&	65.2	$\pm$	0.2	&	28.5	$\pm$	0.3		&	LHS	&	1	&	0.171	$_{-	0.003	} ^{+	0.003	}$   &	0.018	$_{-	0.003	} ^{+	0.006	}$   &	 -					&	 -					&	 -					&	 -						\\
15	&	53433.9	&	91404-01-01-02	&	0.822	$\pm$	0.004	&	87.1	$\pm$	0.2	&	28.9	$\pm$	0.2		&	LHS	&	1	&	0.247	$_{-	0.003	} ^{+	0.002	}$   &	0.047	$_{-	0.002	} ^{+	0.010	}$   &	 -					&	 -					&	 -					&	 -						\\
16	&	53434.7	&	91404-01-01-03	&	0.857	$\pm$	0.005	&	98.8	$\pm$	0.2	&	29.6	$\pm$	0.3		&	LHS	&	1	&	0.311	$_{-	0.005	} ^{+	0.003	}$   &	0.075	$_{-	0.012	} ^{+	0.008	}$   &	 -					&	 -					&	 -					&	 -						\\
17	&	53435.6	&	91404-01-01-01	&	0.807	$\pm$	0.005	&	135.2	$\pm$	0.3	&	29.7	$\pm$	0.3		&	LHS	&	1	&	0.379	$_{-	0.006	} ^{+	0.003	}$   &	0.08	$_{-	0.01	} ^{+	0.01	}$   &	 -					&	 -					&	 -					&	 -						\\
18	&	53436.2	&	91404-01-01-04	&	0.804	$\pm$	0.005	&	157.2	$\pm$	0.4	&	30.4	$\pm$	0.3		&	LHS	&	1	&	0.418	$_{-	0.010	} ^{+	0.004	}$   &	0.072	$_{-	0.012	} ^{+	0.010	}$   &	 -					&	 -					&	 -					&	 -						\\
19	&	53436.4	&	91404-01-01-05	&	0.807	$\pm$	0.006	&	172.7	$\pm$	0.5	&	30.3	$\pm$	0.4		&	LHS	&	1	&	0.469	$_{-	0.006	} ^{+	0.009	}$   &	0.054	$_{-	0.015	} ^{+	0.008	}$   &	 -					&	 -					&	 -					&	 -						\\
20	&	53436.7	&	91702-01-01-00	&	0.800	$\pm$	0.003	&	194.1	$\pm$	0.3	&	30.9	$\pm$	0.1		&	LHS	&	1	&	0.493	$_{-	0.002	} ^{+	0.005	}$   &	0.107	$_{-	0.007	} ^{+	0.008	}$   &	 -					&	 -					&	 -					&	 -						\\
21	&	53437.1	&	91702-01-01-01	&	0.809	$\pm$	0.004	&	207.8	$\pm$	0.4	&	31.1	$\pm$	0.3		&	LHS	&	1	&	0.506	$_{-	0.007	} ^{+	0.006	}$   &	0.12	$_{-	0.01	} ^{+	0.01	}$   &	 -					&	 -					&	 -					&	 -						\\
22	&	53437.1	&	91702-01-01-02	&	0.806	$\pm$	0.005	&	209.0	$\pm$	0.5	&	31.4	$\pm$	0.3		&	LHS	&	1	&	0.520	$_{-	0.004	} ^{+	0.005	}$   &	0.070	$_{-	0.007	} ^{+	0.018	}$   &	 -					&	 -					&	 -					&	 -						\\
23	&	53438.1	&	91702-01-01-03	&	0.764	$\pm$	0.003	&	315.7	$\pm$	0.6	&	31.5	$\pm$	0.2		&	LHS	&	1	&	0.890	$_{-	0.006	} ^{+	0.004	}$   &	0.16	$_{-	0.01	} ^{+	0.01	}$   &	 -					&	 -					&	 -					&	 -						\\
24	&	53438.8	&	91702-01-01-04	&	0.748	$\pm$	0.003	&	415.5	$\pm$	0.6	&	32.5	$\pm$	0.2		&	LHS	&	1	&	1.351	$_{-	0.006	} ^{+	0.003	}$   &	0.208	$_{-	0.013	} ^{+	0.011	}$   &	 -					&	 -					&	 -					&	 -						\\
25	&	53439.1	&	91702-01-01-05	&	0.741	$\pm$	0.003	&	458.1	$\pm$	0.8	&	32.8	$\pm$	0.3		&	LHS	&	1	&	1.529	$_{-	0.006	} ^{+	0.005	}$   &	0.23	$_{-	0.01	} ^{+	0.01	}$   &	 -					&	 -					&	 -					&	 -						\\
26	&	53439.6	&	90704-04-01-01	&	0.716	$\pm$	0.003	&	523.6	$\pm$	0.8	&	32.4	$\pm$	0.2		&	LHS	&	1	&	2.043	$_{-	0.005	} ^{+	0.004	}$   &	0.27	$_{-	0.01	} ^{+	0.01	}$   &	 -					&	 -					&	 -					&	 -						\\
27	&	53439.7	&	90704-04-01-00	&	0.704	$\pm$	0.002	&	561.2	$\pm$	0.8	&	31.9	$\pm$	0.2		&	LHS	&	1	&	2.317	$_{-	0.005	} ^{+	0.004	}$   &	0.29	$_{-	0.01	} ^{+	0.01	}$   &	 -					&	 -					&	 -					&	 -						\\
28	&	53440.7	&	91702-01-02-00G	&	0.455	$\pm$	0.001	&	2026.0	$\pm$	3.0	&	9.6	$\pm$	0.0		&	SIMS	&	5	&	 -					&	 -					&	6.66	$_{-	0.03	} ^{+	0.03	}$   &	2.9	$_{-	0.1	} ^{+	0.1	}$   &	 -					&	 -						\\
29	&	53500.8	&	91702-01-52-03	&	0.387	$\pm$	0.001	&	6471.0	$\pm$	10.0	&	7.1	$\pm$	0.1		&	ULS	&	2	&	18.4	$_{-	0.3	} ^{+	0.3	}$   &	3.2	$_{-	0.5	} ^{+	0.6	}$   &	 -					&	 -					&	8.2	$_{-	0.2	} ^{+	0.2	}$   &	10.7	$_{-	0.5	} ^{+	0.5	}$   	\\
30	&	53501.6	&	91702-01-53-01	&	0.374	$\pm$	0.001	&	6024.0	$\pm$	9.5	&	7.8	$\pm$	0.1		&	ULS	&	4	&	 -					&	 -					&	 -					&	 -					&	9.1	$_{-	0.3	} ^{+	0.3	}$   &	13.4	$_{-	0.6	} ^{+	0.6	}$   	\\
31	&	53502.3	&	91702-01-54-00	&	0.391	$\pm$	0.001	&	6261.0	$\pm$	9.6	&	7.4	$\pm$	0.0		&	ULS	&	2	&	18.9	$_{-	0.1	} ^{+	0.1	}$   &	2.5	$_{-	0.2	} ^{+	0.2	}$   &	 -					&	 -					&	7.8	$_{-	0.1	} ^{+	0.1	}$   &	9.0	$_{-	0.2	} ^{+	0.2	}$   	\\
32	&	53502.4	&	91702-01-54-01	&	0.392	$\pm$	0.001	&	7136.0	$\pm$	11.0	&	7.1	$\pm$	0.1		&	ULS	&	2	&	19.0	$_{-	0.1	} ^{+	0.1	}$   &	3.0	$_{-	0.3	} ^{+	0.4	}$   &	 -					&	 -					&	7.7	$_{-	0.2	} ^{+	0.2	}$   &	9.6	$_{-	0.4	} ^{+	0.4	}$   	\\
33	&	53502.5	&	91702-01-54-02	&	0.395	$\pm$	0.001	&	7721.0	$\pm$	11.8	&	7.4	$\pm$	0.1		&	ULS	&	2	&	19.3	$_{-	0.1	} ^{+	0.1	}$   &	2.4	$_{-	0.3	} ^{+	0.3	}$   &	 -					&	 -					&	7.6	$_{-	0.2	} ^{+	0.2	}$   &	9.5	$_{-	0.4	} ^{+	0.3	}$   	\\
34	&	53502.6	&	91702-01-54-03	&	0.365	$\pm$	0.001	&	5844.0	$\pm$	9.2	&	7.9	$\pm$	0.1		&	ULS	&	2	&	14.7	$_{-	0.5	} ^{+	0.5	}$   &	9.6	$_{-	0.8	} ^{+	0.8	}$   &	 -					&	 -					&	6.0	$_{-	0.5	} ^{+	0.4	}$   &	10.0	$_{-	0.9	} ^{+	1.1	}$   	\\
35	&	53503.2	&	91702-01-55-00	&	0.350	$\pm$	0.001	&	6082.0	$\pm$	9.6	&	7.4	$\pm$	0.0		&	ULS	&	2	&	14.8	$_{-	0.2	} ^{+	0.2	}$   &	8.2	$_{-	0.6	} ^{+	0.6	}$   &	 -					&	 -					&	6.1	$_{-	0.2	} ^{+	0.2	}$   &	11.0	$_{-	0.5	} ^{+	0.5	}$   	\\
36	&	53506.3	&	91702-01-56-00G	&	0.416	$\pm$	0.001	&	7934.0	$\pm$	11.9	&	8.1	$\pm$	0.0		&	ULS	&	2	&	18.06	$_{-	0.02	} ^{+	0.02	}$   &	2.2	$_{-	0.1	} ^{+	0.1	}$   &	 -					&	 -					&	6.2	$_{-	0.1	} ^{+	0.1	}$   &	5.9	$_{-	0.3	} ^{+	0.3	}$   	\\
37	&	53507	&	91702-01-58-03	&	0.397	$\pm$	0.001	&	8423.0	$\pm$	13.1	&	6.7	$\pm$	0.1		&	ULS	&	2	&	19.0	$_{-	0.2	} ^{+	0.2	}$   &	4.5	$_{-	0.5	} ^{+	0.5	}$   &	 -					&	 -					&	7.3	$_{-	0.2	} ^{+	0.2	}$   &	5.1	$_{-	0.8	} ^{+	0.9	}$   	\\
38	&	53507	&	91702-01-58-04	&	0.397	$\pm$	0.001	&	7945.0	$\pm$	12.3	&	7.3	$\pm$	0.1		&	ULS	&	2	&	18.6	$_{-	0.1	} ^{+	0.1	}$   &	2.9	$_{-	0.3	} ^{+	0.3	}$   &	 -					&	 -					&	8.1	$_{-	0.2	} ^{+	0.2	}$   &	8.9	$_{-	0.3	} ^{+	0.3	}$   	\\
39	&	53507.1	&	91702-01-58-02	&	0.382	$\pm$	0.001	&	7001.0	$\pm$	10.9	&	6.5	$\pm$	0.1		&	ULS	&	2	&	18.9	$_{-	0.2	} ^{+	0.2	}$   &	3.8	$_{-	0.4	} ^{+	0.5	}$   &	 -					&	 -					&	6.9	$_{-	0.2	} ^{+	0.3	}$   &	8.8	$_{-	0.4	} ^{+	0.7	}$   	\\
40	&	53507.2	&	91702-01-57-00G	&	0.406	$\pm$	0.001	&	7235.0	$\pm$	10.9	&	7.6	$\pm$	0.0		&	ULS	&	2	&	19.03	$_{-	0.08	} ^{+	0.07	}$   &	2.3	$_{-	0.2	} ^{+	0.2	}$   &	 -					&	 -					&	5.8	$_{-	0.2	} ^{+	0.1	}$   &	6.7	$_{-	0.4	} ^{+	0.4	}$   	\\
41	&	53507.7	&	91702-01-59-00	&	0.419	$\pm$	0.001	&	8130.0	$\pm$	12.4	&	7.7	$\pm$	0.1		&	ULS	&	2	&	18.19	$_{-	0.06	} ^{+	0.05	}$   &	1.9	$_{-	0.1	} ^{+	0.1	}$   &	 -					&	 -					&	6.6	$_{-	0.2	} ^{+	0.2	}$   &	7.8	$_{-	0.2	} ^{+	0.2	}$   	\\
42	&	53508.5	&	91702-01-58-00	&	0.533	$\pm$	0.002	&	10950.0	$\pm$	15.2	&	4.7	$\pm$	0.0		&	ULS	&	6	&	18.7	$_{-	0.1	} ^{+	0.1	}$   &	7.8	$_{-	0.5	} ^{+	0.7	}$   &	6.84	$_{-	0.03	} ^{+	0.03	}$   &	1.0	$_{-	0.1	} ^{+	0.1	}$   &	 -					&	 -						\\
43	&	53509.2	&	91702-01-59-02	&	0.428	$\pm$	0.001	&	7937.0	$\pm$	11.8	&	8.3	$\pm$	0.0		&	ULS	&	2	&	17.44	$_{-	0.02	} ^{+	0.02	}$   &	1.62	$_{-	0.05	} ^{+	0.05	}$   &	 -					&	 -					&	5.9	$_{-	0.1	} ^{+	0.1	}$   &	5.4	$_{-	0.4	} ^{+	0.4	}$   	\\
44	&	53509.6	&	91702-01-58-01	&	0.400	$\pm$	0.001	&	6417.0	$\pm$	9.8	&	7.1	$\pm$	0.0		&	ULS	&	2	&	18.39	$_{-	0.05	} ^{+	0.05	}$   &	2.2	$_{-	0.1	} ^{+	0.1	}$   &	 -					&	 -					&	7.2	$_{-	0.1	} ^{+	0.1	}$   &	8.5	$_{-	0.2	} ^{+	0.2	}$   	\\
45	&	53510	&	91702-01-60-02	&	0.390	$\pm$	0.001	&	7001.0	$\pm$	10.9	&	6.5	$\pm$	0.1		&	ULS	&	2	&	19.0	$_{-	0.2	} ^{+	0.2	}$   &	3.4	$_{-	0.5	} ^{+	0.5	}$   &	 -					&	 -					&	6.6	$_{-	0.2	} ^{+	0.2	}$   &	8.9	$_{-	0.6	} ^{+	0.6	}$   	\\
46	&	53510.1	&	91702-01-60-00	&	0.361	$\pm$	0.001	&	5868.0	$\pm$	9.1	&	6.7	$\pm$	0.0		&	ULS	&	2	&	15.4	$_{-	0.1	} ^{+	0.1	}$   &	7.9	$_{-	0.2	} ^{+	0.3	}$   &	 -					&	 -					&	7.2	$_{-	0.1	} ^{+	0.1	}$   &	9.9	$_{-	0.2	} ^{+	0.2	}$   	\\
47	&	53510.3	&	91702-01-60-01	&	0.374	$\pm$	0.001	&	6113.0	$\pm$	9.5	&	6.6	$\pm$	0.1		&	ULS	&	2	&	14.6	$_{-	0.2	} ^{+	0.2	}$   &	8.5	$_{-	0.4	} ^{+	0.5	}$   &	 -					&	 -					&	7.3	$_{-	0.2	} ^{+	0.2	}$   &	8.1	$_{-	0.5	} ^{+	0.6	}$   	\\
48	&	53515.2	&	91702-01-63-01	&	0.370	$\pm$	0.001	&	6021.0	$\pm$	9.3	&	5.1	$\pm$	0.0		&	ULS	&	2	&	16.54	$_{-	0.09	} ^{+	0.09	}$   &	5.1	$_{-	0.3	} ^{+	0.3	}$   &	 -					&	 -					&	7.7	$_{-	0.1	} ^{+	0.1	}$   &	11.2	$_{-	0.4	} ^{+	0.4	}$   	\\
49	&	53515.7	&	91702-01-63-00	&	0.366	$\pm$	0.001	&	6015.0	$\pm$	9.3	&	5.3	$\pm$	0.0		&	ULS	&	2	&	16.77	$_{-	0.09	} ^{+	0.08	}$   &	5.6	$_{-	0.2	} ^{+	0.2	}$   &	 -					&	 -					&	7.7	$_{-	0.1	} ^{+	0.1	}$   &	9.6	$_{-	0.3	} ^{+	0.3	}$   	\\
50	&	53516.2	&	91702-01-64-02	&	0.353	$\pm$	0.001	&	5878.0	$\pm$	9.3	&	4.5	$\pm$	0.0		&	ULS	&	2	&	17.4	$_{-	0.2	} ^{+	0.2	}$   &	3.1	$_{-	0.5	} ^{+	0.6	}$   &	 -					&	 -					&	8.7	$_{-	0.3	} ^{+	0.3	}$   &	13.5	$_{-	1.0	} ^{+	0.9	}$   	\\
51	&	53516.4	&	91702-01-64-03	&	0.351	$\pm$	0.001	&	5192.0	$\pm$	8.2	&	4.8	$\pm$	0.0		&	ULS	&	2	&	17.0	$_{-	0.2	} ^{+	0.1	}$   &	4.2	$_{-	0.4	} ^{+	0.4	}$   &	 -					&	 -					&	8.5	$_{-	0.2	} ^{+	0.2	}$   &	11.4	$_{-	0.6	} ^{+	0.6	}$   	\\
52	&	53516.6	&	91702-01-64-00	&	0.369	$\pm$	0.001	&	6082.0	$\pm$	9.4	&	4.9	$\pm$	0.0		&	ULS	&	2	&	17.03	$_{-	0.07	} ^{+	0.07	}$   &	4.4	$_{-	0.2	} ^{+	0.2	}$   &	 -					&	 -					&	7.4	$_{-	0.1	} ^{+	0.1	}$   &	10.9	$_{-	0.3	} ^{+	0.3	}$   	\\
53	&	53517	&	91702-01-65-00	&	0.372	$\pm$	0.001	&	6249.0	$\pm$	9.7	&	5.0	$\pm$	0.0		&	ULS	&	2	&	16.7	$_{-	0.2	} ^{+	0.2	}$   &	5.5	$_{-	0.6	} ^{+	0.5	}$   &	 -					&	 -					&	7.3	$_{-	0.2	} ^{+	0.2	}$   &	11.0	$_{-	0.6	} ^{+	0.8	}$   	\\
54	&	53517.1	&	91702-01-65-03	&	0.347	$\pm$	0.001	&	5704.0	$\pm$	9.2	&	4.7	$\pm$	0.1		&	ULS	&	2	&	17.5	$_{-	0.3	} ^{+	0.3	}$   &	4.3	$_{-	0.9	} ^{+	1.1	}$   &	 -					&	 -					&	8.0	$_{-	0.5	} ^{+	0.5	}$   &	12	$_{-	2	} ^{+	2	}$   	\\
55	&	53517.7	&	91702-01-65-01	&	0.299	$\pm$	0.001	&	4799.0	$\pm$	8.1	&	4.5	$\pm$	0.1		&	HSS	&	3	&	20.5	$_{-	0.4	} ^{+	0.3	}$   &	2.0	$_{-	0.7	} ^{+	1.0	}$   &	 -					&	 -					&	 -					&	 -						\\
56	&	53518.1	&	91702-01-66-03	&	0.298	$\pm$	0.001	&	4702.0	$\pm$	7.8	&	4.9	$\pm$	0.0		&	HSS	&	3	&	20.3	$_{-	0.6	} ^{+	0.5	}$   &	7	$_{-	2	} ^{+	2	}$   &	 -					&	 -					&	 -					&	 -						\\
57	&	53518.1	&	91702-01-66-05	&	0.298	$\pm$	0.001	&	4813.0	$\pm$	8.1	&	4.1	$\pm$	0.1		&	HSS	&	3	&	21.3	$_{-	0.5	} ^{+	0.4	}$   &	5	$_{-	1	} ^{+	2	}$   &	 -					&	 -					&	 -					&	 -						\\
58	&	53518.2	&	91702-01-65-02	&	0.297	$\pm$	0.001	&	4779.0	$\pm$	8.0	&	4.4	$\pm$	0.1		&	HSS	&	3	&	20.4	$_{-	0.6	} ^{+	0.6	}$   &	7	$_{-	1	} ^{+	2	}$   &	 -					&	 -					&	 -					&	 -						\\
59	&	53518.7	&	91702-01-66-01	&	0.294	$\pm$	0.001	&	4615.0	$\pm$	7.7	&	4.6	$\pm$	0.0		&	HSS	&	3	&	21.3	$_{-	0.4	} ^{+	0.3	}$   &	6	$_{-	1	} ^{+	1	}$   &	 -					&	 -					&	 -					&	 -						\\
60	&	53518.9	&	91702-01-67-01	&	0.311	$\pm$	0.001	&	4725.0	$\pm$	7.7	&	4.9	$\pm$	0.0		&	HSS	&	3	&	19.6	$_{-	0.3	} ^{+	0.3	}$   &	3.2	$_{-	0.9	} ^{+	1.2	}$   &	 -					&	 -					&	 -					&	 -						\\
61	&	53519	&	91702-01-67-02	&	0.315	$\pm$	0.001	&	4919.0	$\pm$	8.0	&	4.4	$\pm$	0.0		&	HSS	&	2	&	19.6	$_{-	0.6	} ^{+	0.5	}$   &	4.6	$_{-	0.8	} ^{+	1.0	}$   &	 -					&	 -					&	6.8	$_{-	0.6	} ^{+	0.6	}$   &	16.1	$_{-	0.9	} ^{+	1.0	}$   	\\
62	&	53519.1	&	91702-01-67-03	&	0.296	$\pm$	0.001	&	4555.0	$\pm$	7.5	&	4.2	$\pm$	0.0		&	HSS	&	3	&	19.8	$_{-	0.6	} ^{+	0.6	}$   &	6	$_{-	2	} ^{+	3	}$   &	 -					&	 -					&	 -					&	 -						\\
63	&	53526	&	91702-01-72-00	&	0.291	$\pm$	0.001	&	3789.0	$\pm$	6.3	&	3.6	$\pm$	0.1		&	HSS	&	4	&	 -					&	 -					&	 -					&	 -					&	8.6	$_{-	0.6	} ^{+	0.6	}$   &	15	$_{-	1	} ^{+	1	}$   	\\
64	&	53528.6	&	91702-01-73-02	&	0.287	$\pm$	0.001	&	3557.0	$\pm$	6.0	&	3.3	$\pm$	0.1		&	HSS	&	4	&	 -					&	 -					&	 -					&	 -					&	8.8	$_{-	0.7	} ^{+	0.7	}$   &	17	$_{-	1	} ^{+	1	}$   	\\
65	&	53531.5	&	91702-01-77-00	&	0.280	$\pm$	0.001	&	3069.0	$\pm$	5.2	&	3.3	$\pm$	0.1		&	HSS	&	4	&	 -					&	 -					&	 -					&	 -					&	9.4	$_{-	0.7	} ^{+	0.7	}$   &	18	$_{-	1	} ^{+	2	}$   	\\
66	&	53532.2	&	91702-01-77-01	&	0.263	$\pm$	0.001	&	2794.0	$\pm$	4.9	&	3.1	$\pm$	0.1		&	HSS	&	4	&	 -					&	 -					&	 -					&	 -					&	8	$_{-	2	} ^{+	2	}$   &	21	$_{-	3	} ^{+	3	}$   	\\
67	&	53574.4	&	91702-01-16-10	&	0.162	$\pm$	0.001	&	1821.0	$\pm$	3.4	&	3.0	$\pm$	0.0		&	HSS	&	3	&	27.4	$_{-	0.2	} ^{+	0.2	}$   &	1.8	$_{-	0.5	} ^{+	0.6	}$   &	 -					&	 -					&	 -					&	 -						\\
68	&	53575.6	&	91702-01-17-10	&	0.172	$\pm$	0.001	&	1868.0	$\pm$	3.4	&	2.7	$\pm$	0.1		&	HSS	&	3	&	27.51	$_{-	0.10	} ^{+	0.13	}$   &	0.7	$_{-	0.2	} ^{+	0.3	}$   &	 -					&	 -					&	 -					&	 -						\\
69	&	53580.4	&	91702-01-21-10	&	0.161	$\pm$	0.001	&	1823.0	$\pm$	3.4	&	2.8	$\pm$	0.1		&	HSS	&	3	&	27.4	$_{-	0.4	} ^{+	0.4	}$   &	3.1	$_{-	0.7	} ^{+	0.9	}$   &	 -					&	 -					&	 -					&	 -						\\
70	&	53583.4	&	91702-01-24-10	&	0.158	$\pm$	0.001	&	1716.0	$\pm$	3.2	&	2.9	$\pm$	0.1		&	HSS	&	3	&	27.3	$_{-	0.1	} ^{+	0.1	}$   &	1.0	$_{-	0.4	} ^{+	0.4	}$   &	 -					&	 -					&	 -					&	 -						\\
71	&	53585.4	&	91702-01-25-11	&	0.157	$\pm$	0.001	&	1678.0	$\pm$	3.2	&	2.9	$\pm$	0.1		&	HSS	&	3	&	26.9	$_{-	0.2	} ^{+	0.2	}$   &	1.7	$_{-	0.4	} ^{+	0.5	}$   &	 -					&	 -					&	 -					&	 -						\\
72	&	53593.3	&	91702-01-32-10	&	0.147	$\pm$	0.001	&	1514.0	$\pm$	2.9	&	2.7	$\pm$	0.1		&	HSS	&	3	&	25.6	$_{-	0.2	} ^{+	0.4	}$   &	1.4	$_{-	0.4	} ^{+	0.9	}$   &	 -					&	 -					&	 -					&	 -						\\
73	&	53628.2	&	91702-01-76-00	&	0.367	$\pm$	0.001	&	341.5	$\pm$	0.6	&	9.7	$\pm$	0.1		&	HIMS	&	1	&	13.17	$_{-	0.06	} ^{+	0.06	}$   &	1.5	$_{-	0.2	} ^{+	0.2	}$   &	 -					&	 -					&	 -					&	 -						\\
74	&	53628.6	&	91702-01-76-01	&	0.406	$\pm$	0.002	&	332.2	$\pm$	0.8	&	12.1	$\pm$	0.4		&	HIMS	&	1	&	12.74	$_{-	0.10	} ^{+	0.10	}$   &	1.7	$_{-	0.4	} ^{+	0.5	}$   &	 -					&	 -					&	 -					&	 -						\\
75	&	53628.9	&	91702-01-71-03	&	0.500	$\pm$	0.003	&	316.1	$\pm$	0.7	&	16.1	$\pm$	0.3		&	HIMS	&	1	&	9.86	$_{-	0.02	} ^{+	0.02	}$   &	5	$_{-	2	} ^{+	2	}$   &	 -					&	 -					&	 -					&	 -						\\
76	&	53629	&	91702-01-71-04	&	0.487	$\pm$	0.002	&	321.6	$\pm$	0.7	&	15.3	$\pm$	0.3		&	LHS	&	1	&	10.37	$_{-	0.02	} ^{+	0.02	}$   &	0.38	$_{-	0.06	} ^{+	0.07	}$   &	 -					&	 -					&	 -					&	 -						\\
77	&	53629.4	&	91702-01-79-01	&	0.514	$\pm$	0.002	&	310.8	$\pm$	0.6	&	16.7	$\pm$	0.2		&	LHS	&	1	&	9.74	$_{-	0.02	} ^{+	0.01	}$   &	0.56	$_{-	0.04	} ^{+	0.04	}$   &	 -					&	 -					&	 -					&	 -						\\
78	&	53630.5	&	91702-01-79-00	&	0.543	$\pm$	0.002	&	275.0	$\pm$	0.4	&	18.2	$\pm$	0.1		&	LHS	&	1	&	8.93	$_{-	0.01	} ^{+	0.01	}$   &	0.90	$_{-	0.03	} ^{+	0.03	}$   &	 -					&	 -					&	 -					&	 -						\\
79	&	53631.5	&	91702-01-80-00	&	0.584	$\pm$	0.002	&	237.7	$\pm$	0.4	&	20.1	$\pm$	0.1		&	LHS	&	1	&	7.62	$_{-	0.01	} ^{+	0.01	}$   &	0.89	$_{-	0.03	} ^{+	0.03	}$   &	 -					&	 -					&	 -					&	 -						\\
80	&	53632.5	&	91702-01-80-01	&	0.666	$\pm$	0.002	&	194.8	$\pm$	0.3	&	23.8	$\pm$	0.1		&	LHS	&	1	&	4.742	$_{-	0.009	} ^{+	0.008	}$   &	0.62	$_{-	0.03	} ^{+	0.03	}$   &	 -					&	 -					&	 -					&	 -						\\
81	&	53633.5	&	91702-01-81-00	&	0.735	$\pm$	0.004	&	148.4	$\pm$	0.3	&	25.0	$\pm$	0.1		&	LHS	&	1	&	2.24	$_{-	0.02	} ^{+	0.01	}$   &	0.58	$_{-	0.05	} ^{+	0.05	}$   &	 -					&	 -					&	 -					&	 -						\\
82	&	53634.1	&	91702-01-80-02	&	0.749	$\pm$	0.006	&	126.2	$\pm$	0.4	&	25.2	$\pm$	0.3		&	LHS	&	1	&	1.45	$_{-	0.02	} ^{+	0.02	}$   &	0.30	$_{-	0.05	} ^{+	0.05	}$   &	 -					&	 -					&	 -					&	 -						\\
83	&	53634.3	&	91702-01-81-01	&	0.757	$\pm$	0.004	&	119.3	$\pm$	0.2	&	24.9	$\pm$	0.2		&	LHS	&	1	&	1.318	$_{-	0.011	} ^{+	0.006	}$   &	0.23	$_{-	0.03	} ^{+	0.02	}$   &	 -					&	 -					&	 -					&	 -						\\
84	&	53635.5	&	91702-01-81-02	&	0.781	$\pm$	0.004	&	84.7	$\pm$	0.2	&	25.2	$\pm$	0.2		&	LHS	&	1	&	0.592	$_{-	0.005	} ^{+	0.004	}$   &	0.06	$_{-	0.01	} ^{+	0.01	}$   &	 -					&	 -					&	 -					&	 -						\\
85	&	53636.2	&	91702-01-87-03	&	0.786	$\pm$	0.005	&	71.1	$\pm$	0.2	&	25.3	$\pm$	0.4		&	LHS	&	1	&	0.45	$_{-	0.01	} ^{+	0.01	}$   &	0.07	$_{-	0.01	} ^{+	0.04	}$   &	 -					&	 -					&	 -					&	 -						\\
86	&	53636.5	&	91702-01-82-00	&	0.865	$\pm$	0.005	&	58.1	$\pm$	0.2	&	24.4	$\pm$	0.5		&	LHS	&	1	&	0.421	$_{-	0.009	} ^{+	0.009	}$   &	0.10	$_{-	0.01	} ^{+	0.04	}$   &	 -					&	 -					&	 -					&	 -						\\
87	&	53637.2	&	91704-01-01-00	&	0.787	$\pm$	0.006	&	58.9	$\pm$	0.2	&	25.4	$\pm$	0.4		&	LHS	&	1	&	0.327	$_{-	0.004	} ^{+	0.008	}$   &	0.06	$_{-	0.01	} ^{+	0.02	}$   &	 -					&	 -					&	 -					&	 -						\\
88	&	53637.2	&	91704-01-01-01	&	0.791	$\pm$	0.004	&	57.5	$\pm$	0.1	&	25.3	$\pm$	0.2		&	LHS	&	1	&	0.315	$_{-	0.002	} ^{+	0.003	}$   &	0.051	$_{-	0.004	} ^{+	0.006	}$   &	 -					&	 -					&	 -					&	 -						\\
89	&	53637.5	&	91704-01-01-02	&	0.796	$\pm$	0.006	&	56.0	$\pm$	0.2	&	24.7	$\pm$	0.4		&	LHS	&	1	&	0.303	$_{-	0.008	} ^{+	0.003	}$   &	0.043	$_{-	0.009	} ^{+	0.024	}$   &	 -					&	 -					&	 -					&	 -						\\
90	&	53638.4	&	91702-01-86-00	&	0.788	$\pm$	0.005	&	47.7	$\pm$	0.1	&	25.5	$\pm$	0.4		&	LHS	&	1	&	0.231	$_{-	0.007	} ^{+	0.003	}$   &	0.063	$_{-	0.011	} ^{+	0.029	}$   &	 -					&	 -					&	 -					&	 -						\\
91	&	53639.1	&	91702-01-86-01	&	0.806	$\pm$	0.007	&	42.2	$\pm$	0.2	&	25.7	$\pm$	0.5		&	LHS	&	1	&	0.208	$_{-	0.003	} ^{+	0.007	}$   &	0.045	$_{-	0.005	} ^{+	0.027	}$   &	 -					&	 -					&	 -					&	 -						\\
92	&	53639.2	&	91702-01-86-04	&	0.821	$\pm$	0.008	&	41.4	$\pm$	0.2	&	26.3	$\pm$	0.7		&	LHS	&	1	&	0.201	$_{-	0.007	} ^{+	0.004	}$   &	0.062	$_{-	0.017	} ^{+	0.023	}$   &	 -					&	 -					&	 -					&	 -						\\
\hline											\end{longtable} 																																							
\end{landscape}																																																		
\end{center} 																																																		
																																																		
\twocolumn

\bibliographystyle{mn2e.bst}
\bibliography{biblio.bib} 
\label{lastpage}
\end{document}